\documentclass[preprintnumbers,prd,twocolumn, showpacs,floatfix,preprintnumbers,
superscriptaddress,nofootinbib]{revtex4}
\usepackage{graphicx}
\usepackage{epsfig}
\usepackage{bm}
\usepackage{amssymb}
\usepackage{float}
\usepackage{amsmath}
\usepackage[caption=false]{subfig}
\usepackage[export]{adjustbox}
\usepackage{dcolumn}
\usepackage{cancel}
\usepackage[colorlinks]{hyperref}
\usepackage[usenames,dvipsnames]{color}
\hypersetup{
     breaklinks=true,
    pdfstartview={FitH},    % fits the width of the page to the window
    colorlinks=true,       % false: boxed links; true: colored links
    linkcolor=blue,          % color of internal links
    citecolor=red,        % color of links to bibliography
    filecolor=magenta,      % color of file links
    urlcolor=blue,           % c kiolor of external links
    anchorcolor=green,      % Color for anchor text
    linktocpage=true
}
\begin{document}
\title{
Quintessence or Phantom : Study of scalar field dark energy models through a general parametrization of the Hubble parameter  
}
\author{Nandan Roy}
\email{nandan.roy@mahidol.ac.th} 
\affiliation{Centre for Theoretical Physics \& Natural Philosophy, Mahidol University, Nakhonsawan Campus, Phayuha Khiri, Nakhonsawan 60130, Thailand}
\author{Sangita Goswami}
\email{vbsangita91@gmail.com}
\affiliation{Department of Physics, Visva-Bharati, Santiniketan -731235, India}
\author{Sudipta Das}
\email{sudipta.das@visva-bharati.ac.in}
\affiliation{Department of Physics, Visva-Bharati, Santiniketan -731235, India}

\pagestyle{myheadings}

\begin{abstract}
In this work we propose a simple general parametrization scheme of the Hubble parameter for the scalar field dark energy models. In our approach it is possible to incorporate both the quintessence and phantom scalar field in a single analytical scheme and write down relevant cosmological parameters which are independent of the nature of the scalar field. A general condition for the phantom barrier crossing has also been obtained. To test this approach, a well behaved parametrization of the normalized Hubble parameter has been considered and a wide variety of observational data like CMB data, Supernovae data, BAO data etc. has been used to constraint the various cosmological parameters. It has been found that data prefer the present value of the equation of state of the dark energy to be in the phantom domain. One interesting outcome of this analysis is that although the current value of the dark energy equation of state is phantom in nature, a phantom crossing of the EOS has taken place in the recent past. We have also carried out the Bayesian model comparison between $\Lambda CDM$ model and the proposed model which indicates that this model is favored by data as compared to $\Lambda CDM$ model.
\end{abstract}
\maketitle
\section{Introduction}

Even after two decades of its discovery, the reason behind  accelerated expansion of the universe ~\cite{SupernovaSearchTeam:1998fmf, SupernovaCosmologyProject:1998vns,  Meszaros:2002np, Planck:2014loa, ahn2012ninth} remains a mystery. Moreover the observational data also suggest that the onset of this accelerated phase has taken place very  recently, at around $z \sim 0.5$ ~\cite{reiss2001zt, LAmendola2003zt}. If we consider the theory of gravity to be general relativity, any proposed candidate behind the accelerated expansion of the universe referred to as \lq\lq dark energy\rq\rq (DE) must have a sufficient negative pressure to counterbalance gravity and drive the accelerated expansion~\cite{padmanabhan2006dark}. Cosmological constant ($\Lambda$), the most successful of all proposals, is however troubled by the challenges coming from both theoretical and observational sides. An alternative proposal is a dynamical DE model. Till now a variety of such DE models have been proposed, such as quintessence, k-essence, phantom, chaplygin gas, tachyon models, holographic DE models and so on \cite{sahni2006reconstructing, bamba2012dark,armendariz2001essentials, caldwell2002phantom, carroll2003can, kamenshchik2001alternative, sen2002tachyon, padmanabhan2002accelerated, copeland2006dynamics, amendola2010dark}. But the origin and nature of DE still remains unknown despite many years of research and remains an open problem.

 \par In the very recent years, cosmological studies have evidenced an additional open problem, namely the $H_0$ tension which arises as a result of discrepancies in the inferred value of $H_0$ from different types of measurements. The CMB Planck collaboration \cite{Planck2020} (including BAO \cite{BAO2017, BAO2011}, BBN \cite{BBN2021} and DES \cite{1DES2018, 2DES2018, krause2017dark}) have estimated the present value of the Hubble parameter to be $H_0 \sim (67.0 - 68.5)$ km/s/Mpc. On the other hand, cosmic distance ladder and time delay measurements reported by SH0ES \cite{Sh0ES2019} and H0LiCOW \cite{H0LiCOW2019} collaborations have estimated $H_0 = (74.03 \pm 1.42)$km/s/Mpc which has been obtained from the Hubble Space Telescope observations of 70 long-period Cepheids in the Large Magellanic Cloud \cite{Sh0ES2019}.  At the beginning the origin of this discrepancy was thought to be from the systematic. Currently the discrepancy has moved to the order of $\simeq 6 \sigma$ forcing us to expand our thinking beyond the $\Lambda$CDM.
 
\par There have been a number of attempts to address these problems (See \cite{H0Bulk2021, H0Review2021, petronikolou2021alleviating} and the references therein), which includes different dynamical dark energy models with 
a dynamical equation of state ($w_{DE} \ne -1)$  for the dark energy (DE) component \cite{benisty2021preference, karwal2021chameleon, cedeno2021tracker}. It has already been shown by a  number of authors that a phantom-like equation of state (EoS) of the dark energy sector can effectively speed up the  acceleration of the universe which results in a higher value of the Hubble parameter $H_0$ \cite{phantom2020, phantom2021, phantom2020vagnozzi}. 
 
 \par Since there is no preferred theoretical model of dark energy (DE) which can perfectly describe all the observed phenomena of the late time dynamics of the universe, there have been attempts to construct theoretical models of DE right from the observational data. A very useful approach in this context is parametrization of DE models in which a functional form of a particular DE parameter, viz, equation of state (EoS) of the dark energy, Hubble parameter etc.  is chosen. Then the evolutions of the cosmological parameters are studied and various cosmological data sets are used to constrain these parameters.

In this work we revisited the quintessence and phantom scalar field models and present a general scheme to write down the relevant cosmological parameters in terms of the normalized Hubble parameter $E(z)={H(z)}/{H_{0}}$, current matter density $\Omega_{m0}$ and the redshift $z$. The interesting aspect of this approach is that the final expressions of the cosmological parameters do not explicitly depend on the nature of the scalar field. We have considered a parametrization of the Hubble parameter $H(z)$, which is one of the most crucial and important parameter for understanding the evolution of the universe. One can, in principle, parametrize the dark energy equation of state (EoS) $w_{DE}$ or the dark energy density $\rho_{DE}$ as well; but the contributions from the dark energy sector eventually enter the dynamics through the evolution of $H(z)$. This is one of the reasons why we chose to parametrize $H(z)$, as all other  cosmological parameters are directly related to it. Till now a number of parametric forms of Hubble parameter have been proposed (see \cite{banerjee2010chameleon, pacif2017reconstruction, pacif2020accelerating} and the references therein) in the context of quintessence scalar field models of DE. 

We have considered a general approach in which the field equations have been written in terms of the switch parameter $\epsilon$ which can represent both the quintessence ($\epsilon = +1$) as well as phantom scalar field models ($\epsilon =-1$). A publicly available version of the Boltzmann code CLASS~\cite{Blas:2011rf,Lesgourgues:2011rg,Lesgourgues:2011rh} has been amended to study the numerical evolution of the model. A detailed cosmological data analysis has been done using the MCMC code Montepython V3.5~\cite{Brinckmann:2018cvx,Audren:2012wb} using various cosmological data sets. We have also compared our model which we name $\phi CDM$ hereafter with the $\Lambda$ Cold Dark Matter ($\Lambda CDM$) model using the concept of Bayes factor.

 \par The present paper is arranged as follows. The basic mathematical formulation  and the dynamics of the scalar field have been presented in section \ref{Equations}. This section also deals with the parametrization of $E(z)$ and derivation of analytical expressions for various cosmological parameters corresponding to this  parametrization. The observational constraints on the parameters of the model and the numerical evolution of the cosmological system have been summarized in sections \ref{sectionnumericalanalysis} and \ref{sectionevolution} respectively. Finally we conclude by presenting our results and findings in section \ref{conclusions}.    

\section{Scalar field dynamics}\label{Equations}
For a spatially flat, homogeneous and isotropic FRW universe filled  with the scalar field, the Einstein's field equations and KG equation for the scalar field are given by (with  $8\pi G=c=1$)
\begin{center}
\begin{eqnarray}
  3H^2=\rho_{m} +\rho_{\phi}=\rho_{m}+\frac{1}{2} \epsilon \dot{\phi^2}+V(\phi),\label{feq1}\\
  2\dot{H} +3H^2= -p_{\phi}= -\frac{1}{2} \epsilon \dot{\phi^2} + V(\phi),\label{feq2}\\
  \epsilon \Ddot{\phi}+3 \epsilon H\dot{\phi}+  \frac{dV}{d\phi} = 0, \label{kg}
\end{eqnarray} 
\end{center}
    where $ H=\frac{\dot{a}}{a} $ is the Hubble parameter, $\rho_{m}$  
is the matter energy density. The index $\epsilon$, which is named as the switch parameter, characterizes the nature of the scalar field $\phi$. If $\epsilon = +1$ the field is quintessence field and if $\epsilon = -1$ the field is phantom in nature.

The matter energy density is represented by $\rho_{m}$, which varies with the scale factor as $\rho_{m}=\rho_{m0}a^{-3}$ where $\rho_{m0}$ is the current value of the matter energy density. The energy density for the scalar field can be expressed as  $\rho_{\phi}=\frac{1}{2} \epsilon \dot{\phi^2} +V(\phi)$ and the pressure component as $p_{\phi}=\frac{1}{2} \epsilon \dot{\phi^2} - V(\phi)$. $V(\phi)$ is the potential associated with the scalar field $\phi$.\\

By simple rearrangement of terms, from equations (\ref{feq1}), (\ref{feq2}) and (\ref{kg}) it is possible to express the derivative of the Hubble parameter ($\dot{H}$) and scalar field potential $V(\phi)$ as,
 \begin{equation}\label{Hdot}
 2\dot{H} = - \frac{\rho_{m0}}{a^3}-\epsilon \dot{\phi^2 }
 \end{equation}
 and
 \begin{equation}\label{Vphi}
  V(\phi) = \dot{H}+3 H^2-\frac{\rho_{m0}}{2 a^3}   
 \end{equation}

 A further rearrangement of equation (\ref{Hdot}) and use of the standard relation $\dot{H}=\frac{1}{2}a\frac{d}{da}(H^2)$ gives
\begin{center}
\begin{equation}\label{phidot2}
 a\frac{d}{da}(H^2) + \frac{\rho_m{}_0}{a^3} = -\epsilon\dot{\phi^2}. 
\end{equation}
\end{center}

Again $\dot{\phi}$ can be expressed as  \begin{equation}\label{dotphi}
    \dot{\phi}=a H\left(\frac{d\phi}{da}\right) .
\end{equation}
Using equations (\ref{phidot2}) and (\ref{dotphi}), the derivative of the scalar field $\phi$ with respect to redshift $z$ can be expressed as

\begin{equation}\label{dphidzE}
  \frac{d\phi}{dz} =\left[\frac{2E\frac{dE}{dz} -3\Omega_m{}_0(1+z)^2}{\epsilon E^2(1+z)}\right]^\frac{1}{2} 
\end{equation}
Here, we introduce the normalized Hubble parameter $E(z)=\frac{H(z)}{H_{0}}$, where $H_{0}$ is the present value of Hubble parameter and $\Omega_m{}_0=\frac{\rho_m{}_0}{3H_{0}^2}$ corresponds to the present value of the matter density parameter. From now onwards, in the text we would denote $E(z)$ by simply $E$ for the convenience of writing the equations.

In a similar fashion it is possible to express the scalar field potential given in equation (\ref{Vphi}) in terms of $E(z), \Omega_{m 0},$ and $ z$ as
\begin{equation}\label{VphiE}
\frac{V(z)}{3H_{0}^2}=-\frac{(1+z)}{3}E\frac{dE}{dz}+E^2-\frac{1}{2}\Omega_m{}_0(1+z)^3
\end{equation}
One can consider $\frac{V(z)}{3H_{0}^2}$ as the effective potential for the scalar field (in units of critical density $\rho_{crit}=3H_{0}^2$ since $8\pi G=1$). It is interesting to note that in the expressions of $V(\phi)$ or equivalently $V(z)$ given in equations (\ref{Vphi}) or (\ref{VphiE}), there is no explicit dependency on the switch parameter $\epsilon$. This behavior is expected as the nature of a scalar field depends on the kinetic term and not on the potential term.\\
It is also possible to express the density parameter for the matter field $(\Omega_{m})$, density parameter for the scalar field $(\Omega_{\phi})$, the equation of state parameter $(w_{\phi}(z))$ and the deceleration parameter $q(z)$ in terms of the normalized Hubble parameter $E(z)$ as
\begin{equation}\label{Omegam1}
 \Omega_m(z)=\frac{\rho_{m}}{3H^2}=\frac{\Omega_{m0}(1+z)^3}{E^2},
\end{equation}
\begin{equation}\label{Omegaphi1}
\Omega_{\phi}(z)=1-\Omega_{m}(z)=1-\frac{\Omega_{m0}(1+z)^3}{E^2},
\end{equation}
\begin{equation}\label{wphi1}
w_{\phi}(z)=\frac{-1-\frac{2 \dot{H}}{3 H^2}}{\Omega_{\phi}}=\frac{\frac{2}{3}(1+z)E\frac{dE}{dz}-E^2}{E^2-\Omega_{m0}(1+z)^3},
\end{equation}
\begin{equation}\label{q1}
q(z)=-1-\frac{\dot{H}}{H^2}=\frac{(1+z)}{E}\frac{dE}{dz}-1.
\end{equation}

Similar to the potential $V(\phi)$, for the above set of equations (\ref{Omegam1}) to (\ref{q1}) also, it is evident that all the relevant cosmological parameters are independent of the switch parameter $\epsilon$ and hence are identical for quintessence as well as phantom fields. However, it deserves mention here this independence is only apparent and the inherent dependency on $\epsilon$ comes through the evolution of the normalized Hubble parameter $E(z)$. Since the evolution of $E(z)$ directly depends on the nature of the scalar field, i.e, on the value of $\epsilon$, the evolution history will obviously  be different for quintessence and phantom fields even if we consider the same form of potential for both.

One very important aspect to note here is that in equation (\ref{dphidzE}), for the term $\frac{d\phi}{dz}$ to be real, two different conditions needs to be satisfied. For the quintessence field $2 E \frac{dE}{dz} - 3 \Omega_{m0} (1+z)^2 > 0$ and for the phantom field $2 E \frac{dE}{dz} - 3 \Omega_{m0} (1+z)^2 < 0$. More interestingly the dark energy EoS given in equation (\ref{wphi1}) can be recast as 
\begin{equation}
w_\phi (z) = -1 +\frac{1+z}{3 {\Omega_\phi} E^2}\left(2 E \frac{dE}{dz} - 3 \Omega_{mo} (1+z)^2\right),
\end{equation}
which is directly related to the nature of the scalar field as mentioned above. As long as the quantity $2 E \frac{dE}{dz} - 3 \Omega_{m0} (1+z)^2$ is positive, the scalar field will behave as a quintessece field. Eventually as the scalar field evolves with time and the quantity $2 E \frac{dE}{dz} - 3 \Omega_{m0} (1+z)^2$ becomes negative, the field will behave as a phantom one. The phantom crossing will occur at a redshift $z=z_{\lambda}$ for which $2 E \frac{dE}{dz} - 3 \Omega_{m0} (1+z)^2 = 0$ and the EoS of the scalar field will coincide with that for the cosmological constant. A general conclusion regarding the phantom barrier crossing for the DE models can be drawn from here. Any dark energy model for which $2 E \frac{dE}{dz} - 3 \Omega_{m0} (1+z)^2 > 0$ for $z>z_{\lambda}$, and $2 E \frac{dE}{dz} - 3 \Omega_{m0} (1+z)^2 < 0$ for $z<z_{\lambda}$ should undergo a phantom barrier crossing.

Conventionally one can consider a particular form of a potential $V(\phi)$ and study the evolution of the quintessence field and phantom field separately. In our current approach one can consider a particular parametrization of $E(z)$ and  reconstruct all the cosmological parameters from it. Though consideration of a particular form of $E(z)$ is equivalent to the consideration of a potential in view of equation (\ref{VphiE}), the advantage of our approach is that one can study both the quintessence and phantom field in a single framework. Moreover recent cosmological data sets can be used to constraint the cosmological parameters together with the EoS of the dark energy which can also help us to determine the nature of the scalar field, quintessence or phantom.

In the present work we have considered the following parametrization of the normalized Hubble parameter $E(z)$ as toy model, in order to reconstruct other cosmological parameters,

\begin{equation}\label{Eparam1}
E(z)=\left[1+pz\left(b+ \frac{z}{c} +\frac{z^2}{d}\right)\right]^\frac{1}{2}
\end{equation} 
 where $b$, $c$, $d$ and $p$ are nonzero real numbers. Hence the Hubble parameter $H(z)$ can be expressed as,

 \begin{equation}\label{H2param1}
H^2(z)= {H_0}^2\left [1+pz\left(b+ \frac{z}{c} +\frac{z^2}{d}\right)\right].
 \end{equation} 
 The functional form chosen in equation (\ref{Eparam1}) (or equivalently in equation (\ref{H2param1})) is completely phenomenological, where $H^2(z)$ happens to be a polynomial in $z$. Also it can be seen from equation (\ref{Eparam1}) that at $z=0$, $H(z)=H_{0}$ and one gets back the present value of the Hubble parameter. Notice that for the choice of $p=0$, the above equation also represents the current value of the Hubble parameter, thus the second term in the above parametrization can be considered as the deviation from the present value of the normalized Hubble parameter. For non zero values of the parameter $p$, at late times when $z << 1$, the deviation is small but for early times, when $z>1$, it can have large deviation depending on the choice of the parameters. This behaviour is expected for any parametrized quantity from the observational point of view.  \\
 
 Now for this particular choice of $E(z)$ given by (\ref{Eparam1}), equations (\ref{Omegam1}), (\ref{Omegaphi1}), (\ref{wphi1}) and (\ref{q1}) take the form 
\begin{equation} \label{eq:om}
\Omega_{m}(z)=\frac{\Omega_{m0}(1+z)^3}{\left[1+pz\left(b+ \frac{z}{c} +\frac{z^2}{d}\right)\right]}
\end{equation}
   \begin{equation}
  \Omega_{\phi}(z)=1-\frac{\Omega_{m0}(1+z)^3}{\left[1+pz\left(b+ \frac{z}{c} +\frac{z^2}{d}\right)\right]}
\end{equation}
  \begin{equation} \label{eq.EoS1}
 w_{\phi}(z) =\frac{p\left(b-2bz+\frac{2z}{c}-\frac{z^2}{c}+\frac{3z^2}{d}\right)-3}{3\left[1+pz\left(b+ \frac{z}{c} +\frac{z^2}{d}\right)-\Omega_m{}_0(1+z)^3\right]}
\end{equation}
and
\begin{equation}\label{qE1}
q(z)=\frac{p\left(b-bz+\frac{2z}{c}+\frac{3z^2}{d}+\frac{z^3}{d}\right)-2}{2\left[1+pz\left(b+ \frac{z}{c} +\frac{z^2}{d}\right)\right]}
 \end{equation}

For the sake of completeness, using equation  (\ref{Eparam1}) in equations (\ref{dphidzE}) and (\ref{VphiE}), we have also provided the expressions for the potential $\phi(z)$ and $V(\phi)$ for this particular choice of $E(z)$.
\begin{equation}
\phi (z)=\int{\left[\frac{p\left(b+\frac{2z}{c}+\frac{3z^2}{d}\right)-3\Omega_{m0}(1+z)^2}{\epsilon (1+z)\left[1+pz\left(b+ \frac{z}{c} +\frac{z^2}{d}\right)\right]}\right]^\frac{1}{2}} dz
\end{equation}
and
\begin{equation}\label{eq:vphi}
\begin{split}
\frac{V(z)}{3H_{0}^2}=-\frac{p(1+z)\left(b+\frac{2z}{c}+\frac{3z^2}{d}\right)}{6}+\\
\left[(1+pz\left(b+ \frac{z}{c} +\frac{z^2}{d}\right)\right]-\frac{1}{2}\Omega_{m0}(1+z)^3
\end{split}
\end{equation}
  The expression for the scalar field EoS $w_\phi$, given in equation (\ref{eq.EoS1}), can be equivalently written in a more compact form in terms of the scale factor $(\frac{a}{a_0})$ as
\begin{equation} \label{eq.EoS2}
     {w}_{\phi}=\frac{3 w_0\left(\frac{a}{a_{0}}\right)^{3}+2 w_1\left(\frac{a}{a_{0}}\right)^{2}+w_3\left(\frac{a}{a_{0}}\right)}{3\left[w_2 -w_0\left(\frac{a}{a_{0}}\right)^{3}-w_1\left(\frac{a}{a_{0}}\right)^{2}-w_3\left(\frac{a}{a_{0}}\right)\right]},
\end{equation}
   where $w_0,~ w_1,~ w_2,~w_3$ are new set of parameters which are related to our old set of parameters $p, b, c,d$ in the following way :
 $$w_0=[p(b c d-d+c)-c d],~~ w_1=p[2 d-b c d-3 c],$$ $$w_3=p(3 c-d)     ~~{\mathrm{and}}~~ w_2=p c-\Omega_{m 0} c d~.$$

All others cosmological parameters can also be written in terms of the new set of parameters $w_0,~ w_1,~ w_2,~w_3$ (please see Appendix~\ref{appen:parameters} for details). From equation (\ref{eq.EoS2}), the present value of the EoS of the dark energy $w_{\phi 0}$ can be expressed as $$w_{\phi 0} = -1 + \delta(w_0, w_1, w_2, w_3),$$ where $\delta(w_0, w_1, w_3) = \frac{w_1 - 3w_2 + 2w_3}{3(w_0 + w_1 + w_3 - w_2)}$ can be considered as the deviation from the pure $\Lambda$CDM model at present. Hence, depending on the values of the model parameters ($w_0, w_1, w_2, w_3$) or equivalently ($b, c, d, p$ and $\Omega_{m 0}$), the present value of the EoS will be either in the quintessence region or in the phantom region. 
 
 This general setup gives us the advantage to study both the quintessence and phantom field together. It would be interesting to investigate the dynamics of the dark energy sector considering the above mentioned parametrization of the EOS of the dark energy, or equivalently the scalar field models, against the current cosmological data. 
 
 \section{Numerical Investigation and Observational Constraint} \label{sec:numerical} 
 Our aim here is to study the numerical evolution of our model and find the observational constraints on the cosmological parameters by comparing it to the observational data sets.  In order to do so  we have used the expression of $w_\phi$ given in equation (\ref{eq.EoS2}) and amended a public version of the CLASS Boltzmann code to include it in the  dark energy sector as a fluid. From now onwards, we will name this scalar field model as $\phi$CDM model. It has been shown in \cite{Roy:2018nce, Roy:2018eug} that the late time dynamics of the scalar field as the component of the dark energy is almost independent of the form of the potentials; it is the EoS of the scalar field which affects the dynamics. Thus it is expected that, consideration of the EoS of the scalar field as a fluid in the CLASS code will resemble the same dynamics as the scalar field itself. In appendix \ref{appen:comparison}, we have presented a comparison between the analytical solutions for different cosmological parameters obtained in this work and the corresponding numerical solutions obtained from CLASS code. The MCMC code Montepython3.5 \cite{Brinckmann:2018cvx} has been used to estimate the relevant cosmological parameters. 
 
 \subsection{Observational Constraints}\label{sectionnumericalanalysis}
Cosmological data sets which we have used for the analysis are as follows: Pantheon\cite{Scolnic:2017caz}, BAO (BOSS DR12~\cite{Alam_2017}, 6dFGS~\cite{Beutler_2011}, eBOSS DR14 (Lya)~\cite{Cuceu_2019} and WiggleZ~\cite{Kazin_2014}). We have also included (SDSS LRG DR7~\cite{Ross_2015}, SDSS LRG DR4~\cite{Tegmark_2006} and WiggleZ~\cite{Kazin_2014}) which are the observations related to  cluster counts. A SH0ES prior together with  the compressed Planck likelihood has been imposed. The compressed Planck likelihood has more or less the same constraining ability as the full Planck likelihood. It has been constructed by considering only three parameters, the baryon physical density $\omega_b = \Omega_b h^2$ and the two shift parameters $\theta_\ast = r_s(z_{dec})/D_A(z_{dec})$ and $\mathcal{R} = \sqrt{\Omega_M H^2_0} D_A(z_{dec})$, where $z_{dec}$ is the redshift at decoupling and $D_A$ is the comoving angular diameter distance. For more details on the compressed Planck likelihood please see  Appendix~A of \cite{Arendse_2020}.

We have made the choice of flat priors on the base cosmological parameters as follows: the baryon density $100 \omega_b = [1.9,2.5]$; cold dark matter density $\omega_{cdm} = [0.095,0.145]$; Hubble parameter $H_0=[60,80] \, \mathrm{km \, s^{-1} \, Mpc^{-1}}$.
  
   A wide range of flat prior on the model parameters $w_0,w_1,w_3$: $w_0=[0,300]$, $w_1=[-50,50]$ and $w_3=[-50,50]$ has been considered. We have fixed the model parameter $w_2$ to be zero. This choice has been considered from the point of view of the stability of the CLASS code and also to minimize the shooting failure. The choice of $w_2= c(p - \Omega_{m0}~d)=0$ establishes a relation between $p$, $\Omega_{m0}$ and $d$ as $\Omega_{m0}=\frac{p}{d}$, which does not put any tight constraint on other model parameters $w_0,~w_1$ and $w_3$; there will be always room for suitable choice of parameters which can lead to a viable cosmological model. 

\begin{figure*}[htp!]
\includegraphics[width=\textwidth]{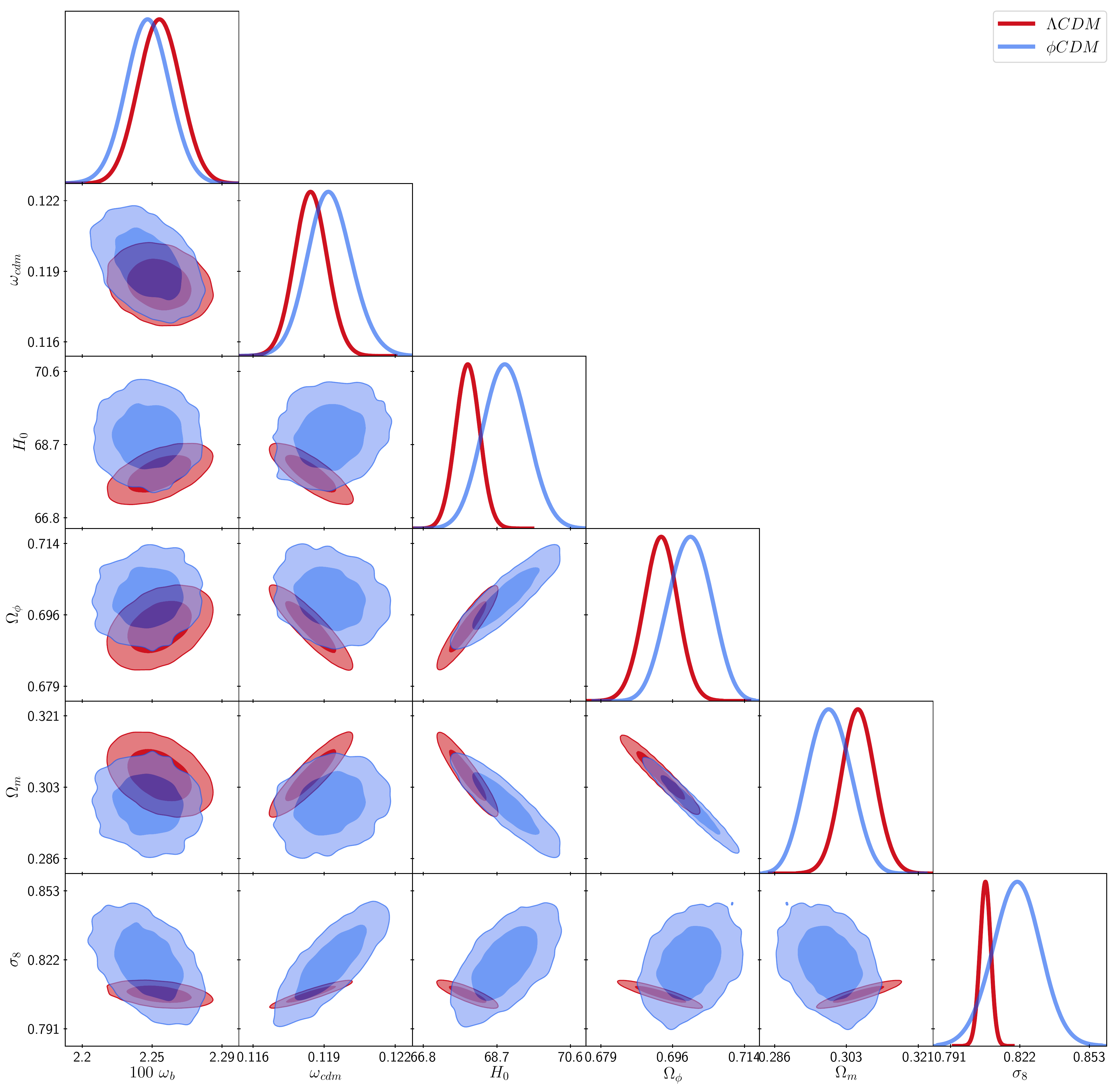}
\caption{\label{fig:cosmo} Triangular plot of 2D and 1D posterior distribution of the cosmological parameters $100 w_1$, $w_{cdm}$,$H_0$, $\Omega_m$ and $\sigma_8$ using various dataset mentioned in section \ref{sectionnumericalanalysis}.}
\end{figure*}

In Fig \ref{fig:cosmo}, we have shown the triangular plot which shows the 2D and 1D posterior distribution of the cosmological parameters $100 w_b, w_{cdm}, H_0, \Omega_m$ and $\sigma_8$. A comparison has been made  with the $\Lambda CDM$ case represented by red plots in the figure. The constraints on the various cosmological parameters for the $\Lambda CDM$ and the $\phi CDM$ models have been enlisted in Table \ref{tab:parameters}. Although there is a slight increment in the value of $H_0$ for the $\phi CDM$  model, but it is far from solving the Hubble tension. A similar result of slight increment of the $H_0$ parameter from $\Lambda CDM$ value for the scalar field models has been reported in \cite{Vazquez:2020ani,LinaresCedeno:2021aqk}. Both these works analyze a large class of potentials for scalar field models and reported its inability to solve the $H_0$ tension.

\begin{figure}[htp!]
\includegraphics[width=\columnwidth]{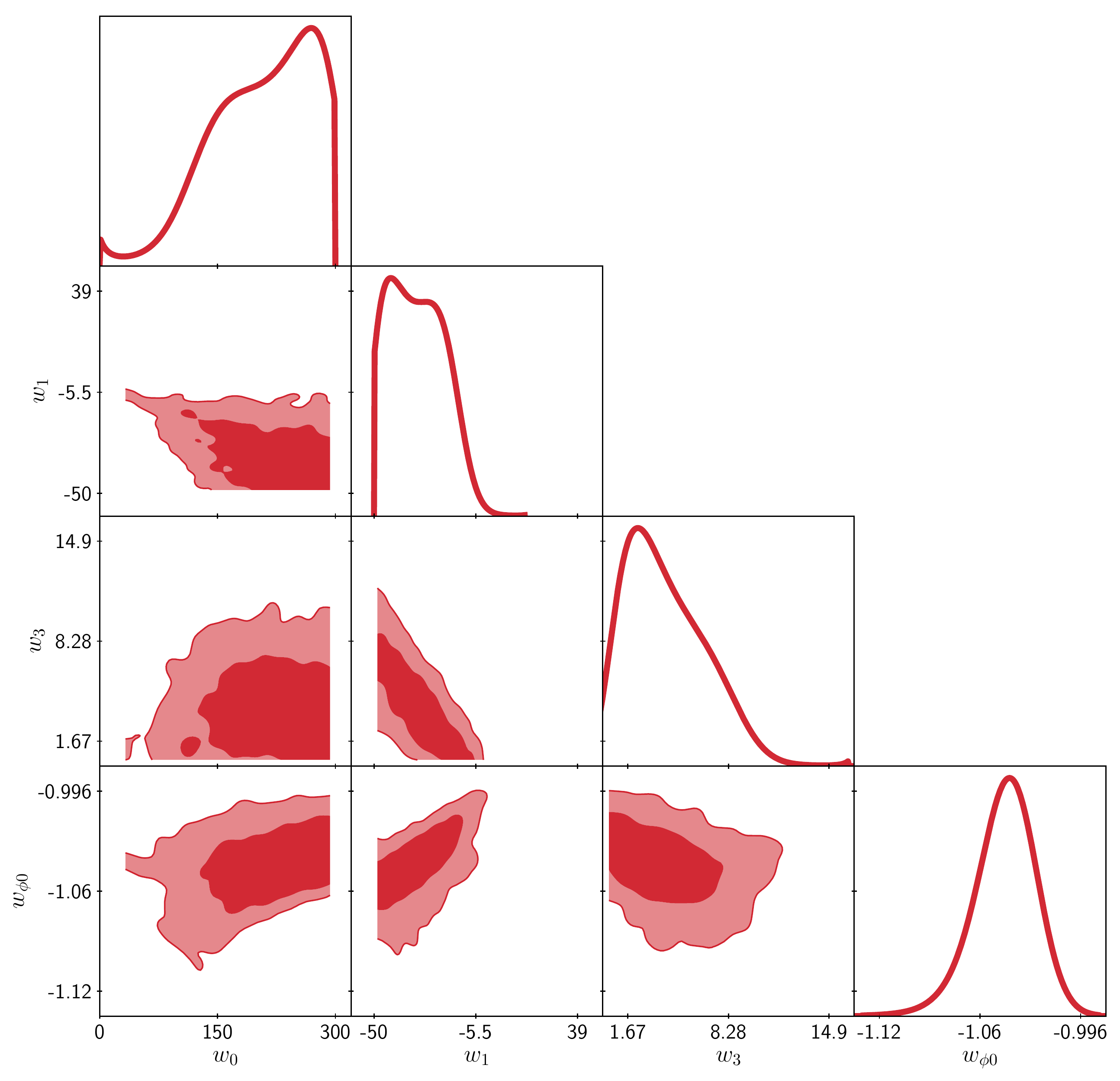}
\caption{\label{fig:model} Triangular plot showing observational constraints on the model parameters $w_0, w_1, w_3$ and the EoS of the scalar field $w_{\phi 0}$.}
\end{figure}

\begin{figure}[htp!]
\includegraphics[width=\columnwidth]{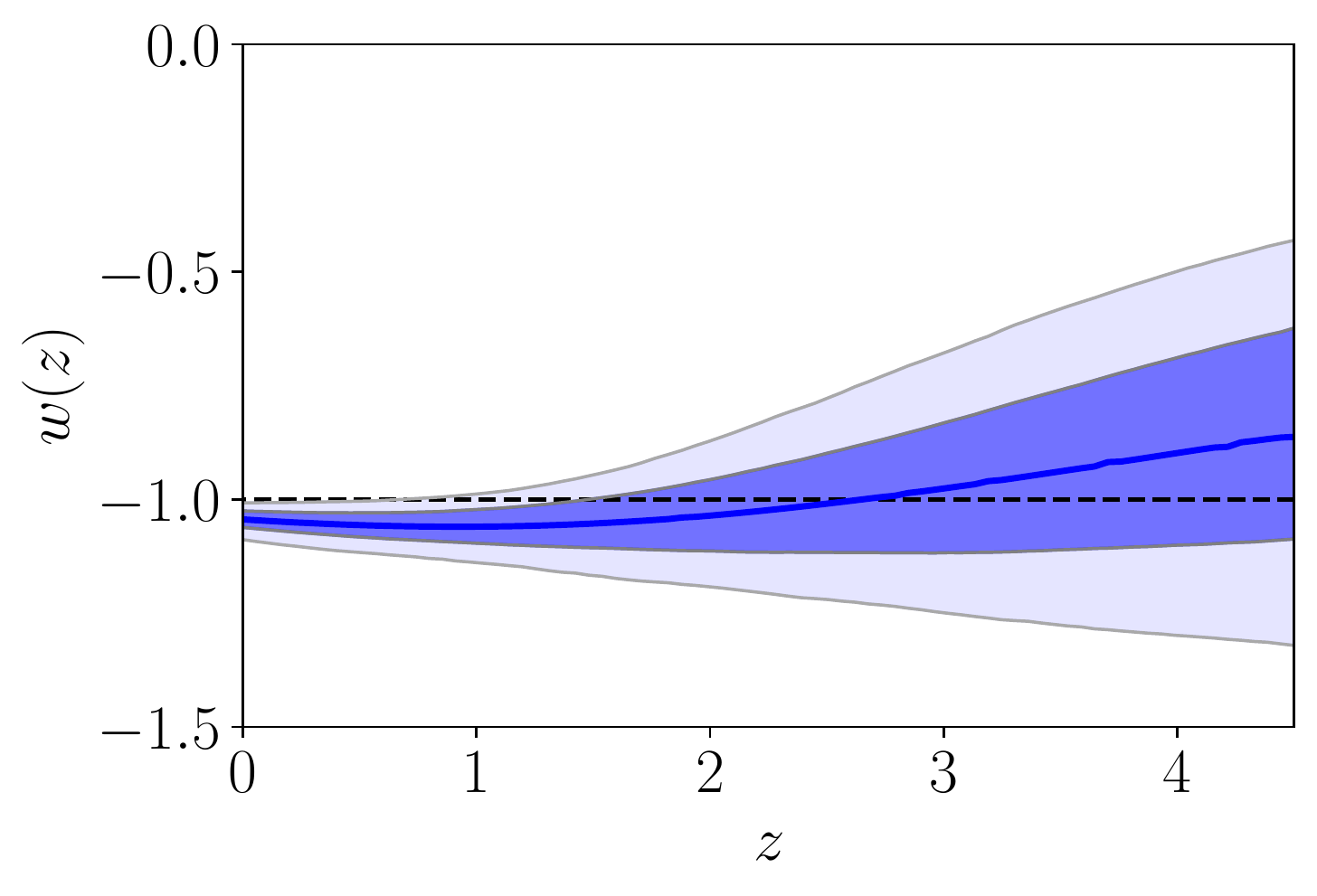}
\caption{\label{fig:w_phi} This figure shows the posterior probability $Pr(w \vline z)$ of the dark energy EOS $w_\phi$ against $z$. The deep blue region represents $1\sigma (68\%)$ confidence contour level and the light blue regions represent the $2\sigma (95\%)$ confidence contour level.} 
\end{figure}

The observational constraints on the model parameters including the dark energy EoS are shown in Fig \ref{fig:model}. The best fit value of the EoS of the dark energy is found to be $w_{DE} =-1.04^{+0.0204}_{-0.0166}$ which is in the phantom region. In Fig \ref{fig:w_phi}, we have plotted the posterior probability $Pr(w \vline z)$ of the dark energy EOS $w_\phi$ against $z$. The deeper blue region enclosed by the solid lines represents $1\sigma (68\%)$ contour level and the lighter blue region represents $2\sigma (95\%)$ contour level. This plot suggests a phantom crossing of the universe around $z=2.7$ which occurs following the conditions mentioned in section \ref{Equations}. 

\begin{figure}[htp!]
\includegraphics[width=\columnwidth]{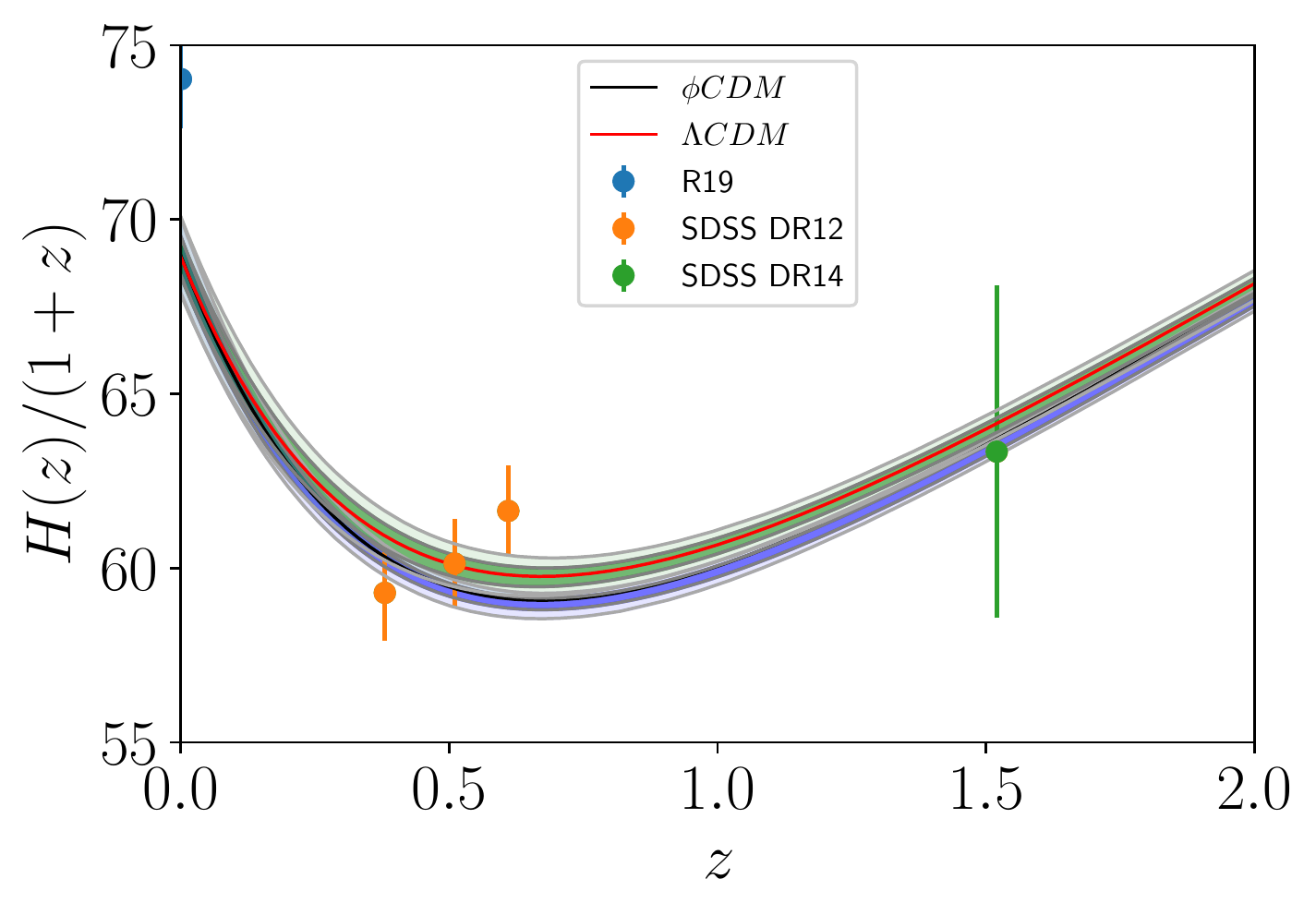}
\caption{\label{eq:H} The posterior probability $Pr(H \vline z)$ of the expansion rate of the universe $H(z)/(1+z)$ against $z$. The dark gray region represents $1\sigma (68\%)$ contour level and the light gray regions represent $2\sigma (95\%)$ contour level.} 
\end{figure}

\begin{figure}[htp!]
\includegraphics[width=\columnwidth]{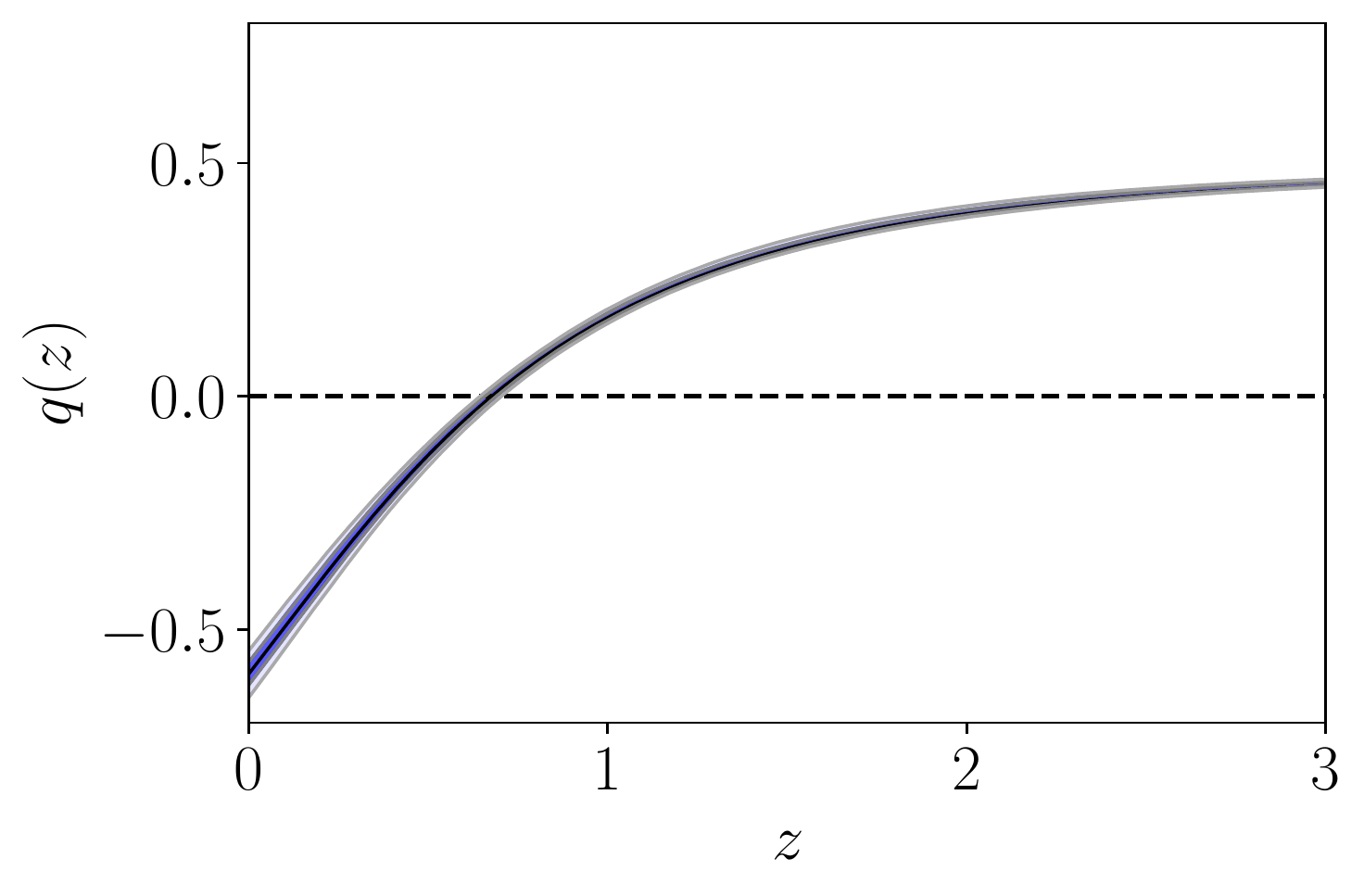}
\caption{\label{fig:q} This figure shows the posterior probability $Pr(q \vline z)$ of the deceleration parameter $q(z)$ against $z$. The deep blue region represents $1\sigma (68\%)$ contour level and the light blue regions represents $2\sigma (95\%)$ contour level.} 
\end{figure}

In Fig \ref{eq:H}, we have shown the posterior probability of the expansion rate of the universe $H(z)/(1+z)$ with respect to $z$. For comparison we have plotted the observational data from Sh0ES \cite{Riess:2019cxk} and BAO observations \cite{BOSS:2016wmc,Zarrouk:2018vwy,Blomqvist:2019rah,deSainteAgathe:2019voe}. The blue line corresponds to the $\phi CDM$ model and the red line corresponds to the $\Lambda CDM $ model. The deeper and lighter blue  and green regions represent the $1\sigma (68\%)$ and $2\sigma (95\%)$
contour level for the $\phi CDM$ and $\Lambda CDM$ model respectively. A similar plot for the posterior probability of the  deceleration parameter $q(z)$ is given in Fig \ref{fig:q} with $1\sigma$ and $2\sigma$ confidence level. A flip in the signature of the deceleration parameter is observed which indicates a transition from the decelerated phase to an accelerated phase which is a must for the unhindered structure formation of the universe.

To compare the $\Lambda CDM $ model and our proposed $\phi CDM$ model, we have used $\chi^2_{min}$ value difference between the models, $\Delta \chi^2_{min} = \chi^2_\phi - \chi^2_\Lambda = -5$ (see Table \ref{tab:parameters}). This indicates that the proposed $\phi CDM $ model gives a better fit to the data as compared to the $\Lambda CDM$ model. We have also computed the Bayesian evidence for model selection to be more certain about the preference on the model from data. Bayes factor is used to penalize the complex models involving many parameters and hence to avoid overfitting. To compare the models we have computed $\ln B_{\phi \Lambda} = \ln \mathcal{Z}_\phi - \ln \mathcal{Z}_\Lambda$, where $\mathcal{Z}$ is the Bayesian evidence. According to Jeffreys’ scale,  the strength of preference on $\phi CDM$ over $\Lambda CDM$ will depend on the value of $\ln B_{\phi \Lambda}$. If $\ln B_{\phi \Lambda} <1$, the preference is negative and if $\ln B_{\phi \Lambda}>1; >2.5; >5.0$ the preference is positive, moderate and strong respectively. For more details, one can refer to \cite{Trotta:2005ar}. The calculation of the Bayes factor is numerically challenging; for this, we have used the publicly available code \textsc{MCEvidence}~\cite{Heavens:2017afc} which can calculate the Bayes factor directly from the MCMC chains generated by the  \textsc{Monte Python}. As listed in Table \ref{tab:parameters}, we have obtained $\ln B_{\phi \Lambda} = + 2.005$  which indicates positive preference on the $\phi CDM$ model compared to the $\Lambda CDM$ model.

\begin{table}[h]
\caption{\label{tab:parameters} Best fit values of different cosmological parameters for the $\Lambda CDM$ and $\phi CDM $ models. The posteriors can be seen from the triangular plots provided in Fig \ref{fig:cosmo} and Fig \ref{fig:model}}
\begin{tabular}{|l|c|c|c|c|} 
 \hline \hline
Parameter & $\Lambda CDM$ & $\phi CDM$ \\ \hline 
$100~\omega{}_{b }$ &$2.25^{+0.013}_{-0.0129}$ & $2.24^{+0.0146}_{-0.0128}$  \\ \hline
$\omega{}_{cdm }$ &$0.118^{+0.00069}_{-0.000705}$ & $0.119^{+0.0096}_{-0.0098}$  \\ \hline
$H_0$ &$68^{+0.326}_{-0.319}$ & $68.9^{+0.573}_{-0.602}$  \\ \hline
$\Omega_{ DE }$ &$0.694^{+0.00426}_{-0.00406}$ & $0.701^{+0.00478}_{-0.00516}$ \\ \hline
$\Omega_{m }$ &$0.306^{+0.00406}_{-0.00426}$ & $0.299^{+0.00516}_{-0.00478}$ \\ \hline  
$\sigma_{8}$ &$0.807^{+0.00255}_{-0.00254}$ & $0.821^{+0.011}_{-0.010}$ \\ \hline
$w_{DE }$ &$-1$ & $-1.04^{+0.0204}_{-0.0166}$  \\ \hline
$w_{0 }$ &$-$ & $>96.97$  \\ \hline
$w_{1 }$ &$-$ & $<-10.03$  \\ \hline
$w_{3 }$ &$-$ & $4.25^{+1.65}_{-3.46}$  \\ \hline
$\Delta \chi_{min}^2$ &$0$ & $-5$ \\ \hline
$\ln{B_{\phi \lambda}}$ &$0$ & $+4.225$ \\ \hline
\hline 
\end{tabular} \\ 
\end{table}

\subsection{Numerical Evolution}\label{sectionevolution}

\begin{figure}[h!]
\includegraphics[width=0.49\textwidth]{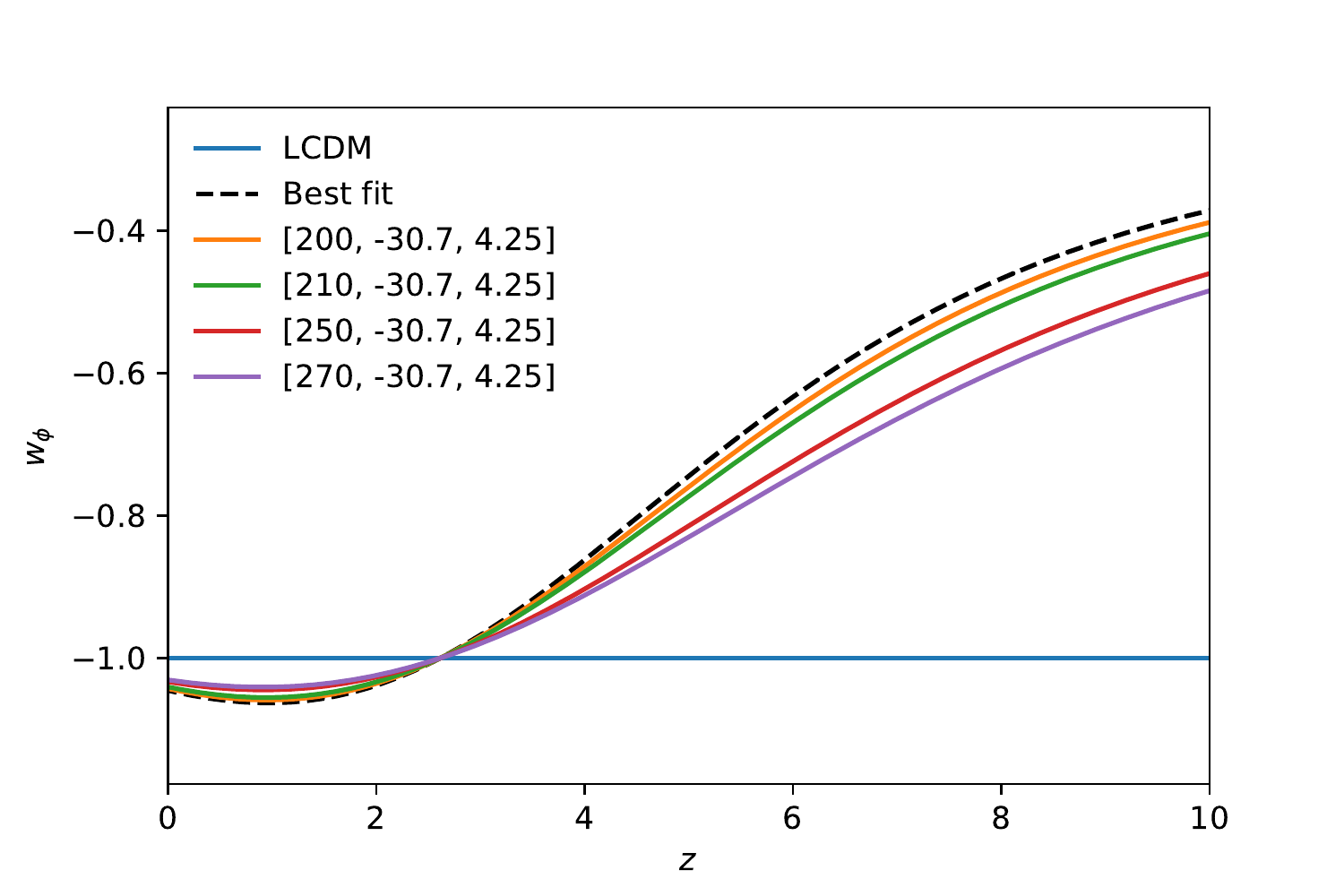}
\includegraphics[width=0.49\textwidth]{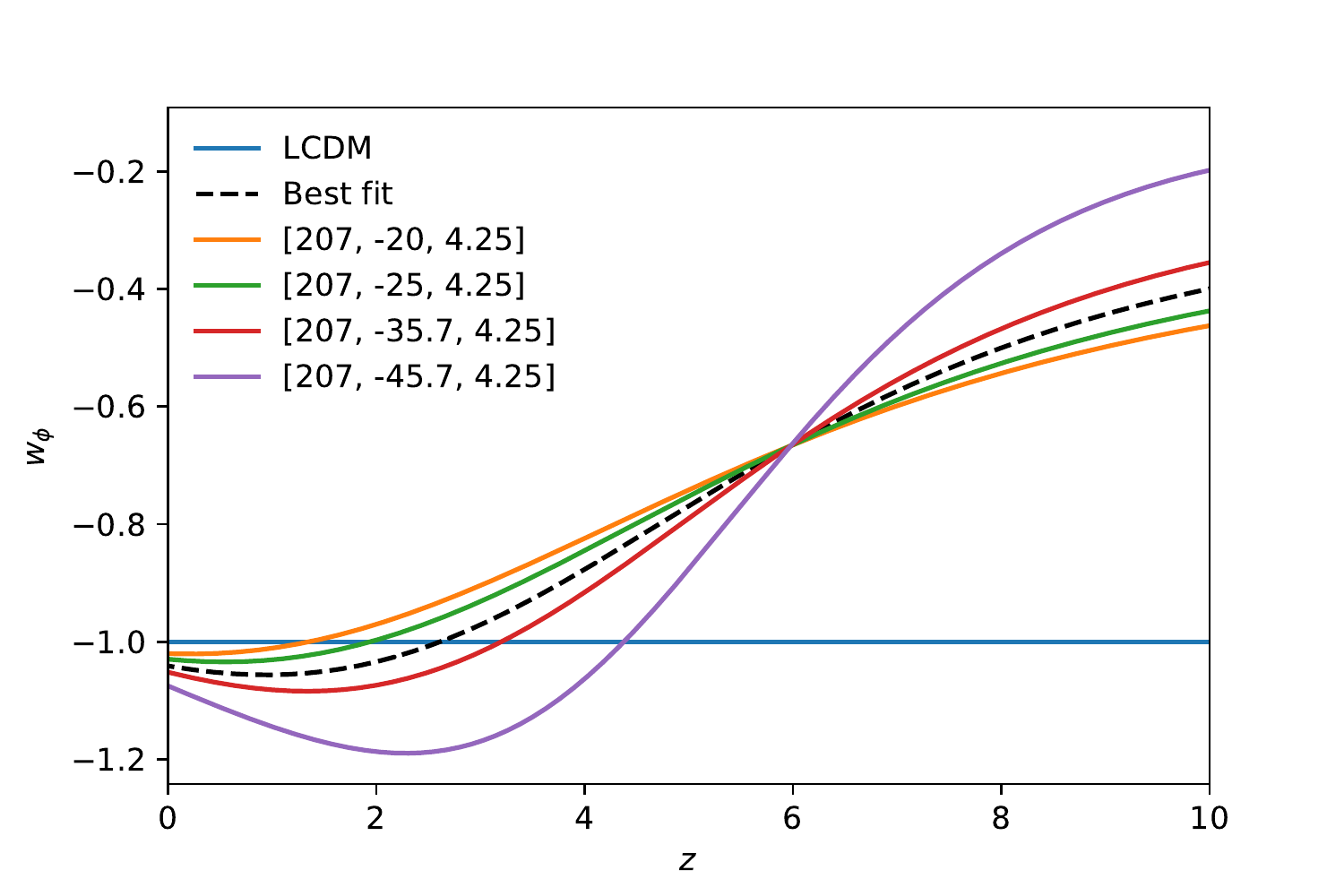}
\includegraphics[width=0.49\textwidth]{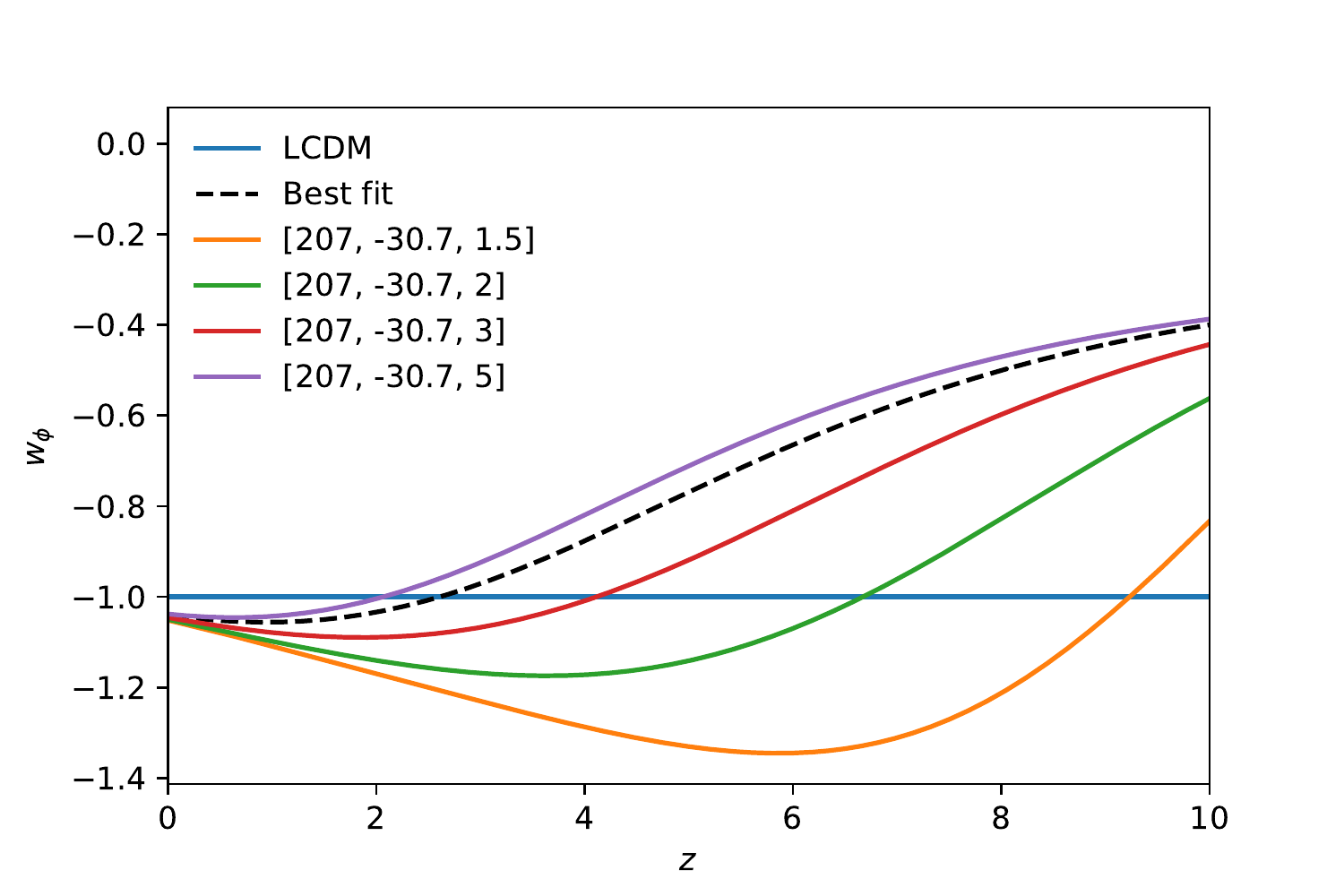}
\caption{\label{fig:EoS} The evolution of the EoS ($w_\phi$) for the $\phi CDM$ model for different choices of the model parameters $(w_0, w_1, w_3)$ against the redshift $z$. The offset values refer to the respective choices of the model parameters. The solid blue line represents the $\Lambda CDM $ model and the black dashed line  represents the evolution of the EoS corresponding to the best fit values obtained from MCMC analysis. The top panel shows the evolution when we are varying only $w_0$ while keeping the parameters $w_1, w_3$ fixed. The middle and bottom panels show the evolution when $w_1$ and $w_3$  are evolving keeping the other two respective parameters fixed. The blue horizontal line corresponds to the $\Lambda$CDM case $w_\Lambda =-1$.}
\end{figure}

We have considered the observational constraints obtained on the parameters $w_0,~w_1,~  w_3$ in the previous section to investigate the numerical evolution of the system.  Fig \ref{fig:EoS} shows the evolution of the EoS parameter of the scalar field component $w_\phi$ with respect to redshift $z$. To understand the effect of the model parameters $w_0,w_1, w_3$ on the cosmological parameters, we consider to vary one parameter keeping others fixed to the best fit values obtained from the MCMC analysis. The black dashed lines in all the plots correspond to the best fit values of the model parameters $[w_0, w_1, w_3:207,-30.7,4.25]$. In the top panel of Fig \ref{fig:EoS}, we vary $w_0$ while keeping the parameters $w_1$ and $w_3$ fixed to the best-fit value. In the middle and bottom panels, we vary $w_1$ and $w_3$ respectively keeping the other two parameters fixed. Our choice of parameters lies well within the posteriors obtained by the MCMC analysis in the previous section. From Fig \ref{fig:EoS} one can see that the current value of EoS of the scalar field  lies in the phantom region ($w_\phi < -1$) and the universe should have undergone phantom crossing from the quintessence to phantom (see the best fit curve, black dotted line). From the top panel it is apparent that the effect of $w_0$ on the EoS parameter is not very  significant at later epochs and also the redshift at which phantom crossing occurs is insensitive to the value of $w_0$. The effect of the parameter $w_0$ is only apparent at the early epochs of the evolution of the universe. However, the present value of the EoS parameter as well as the evolution of the EoS is quite sensitive to both the parameters $w_1, w_3$. This finding is in agreement with the results shown in Fig \ref{fig:model} obtained by MCMC analysis. It has been observed that the evolution of the EoS parameter will be effected only when the variation of $w_1$ is of the order of 10 or more. The $w_0, w_1$ parameters appear to be unconstrained with a lower cutoff of $w_0 > 96.97$ (approximately) and a upper cutoff $w_1<-5.5$ (approximately) respectively. On the other hand $w_3$ is tightly constrained with allowed values in the range $4.25^{+1.65}_{-3.46}$.

\begin{figure}[h]
\includegraphics[width=0.48\textwidth]{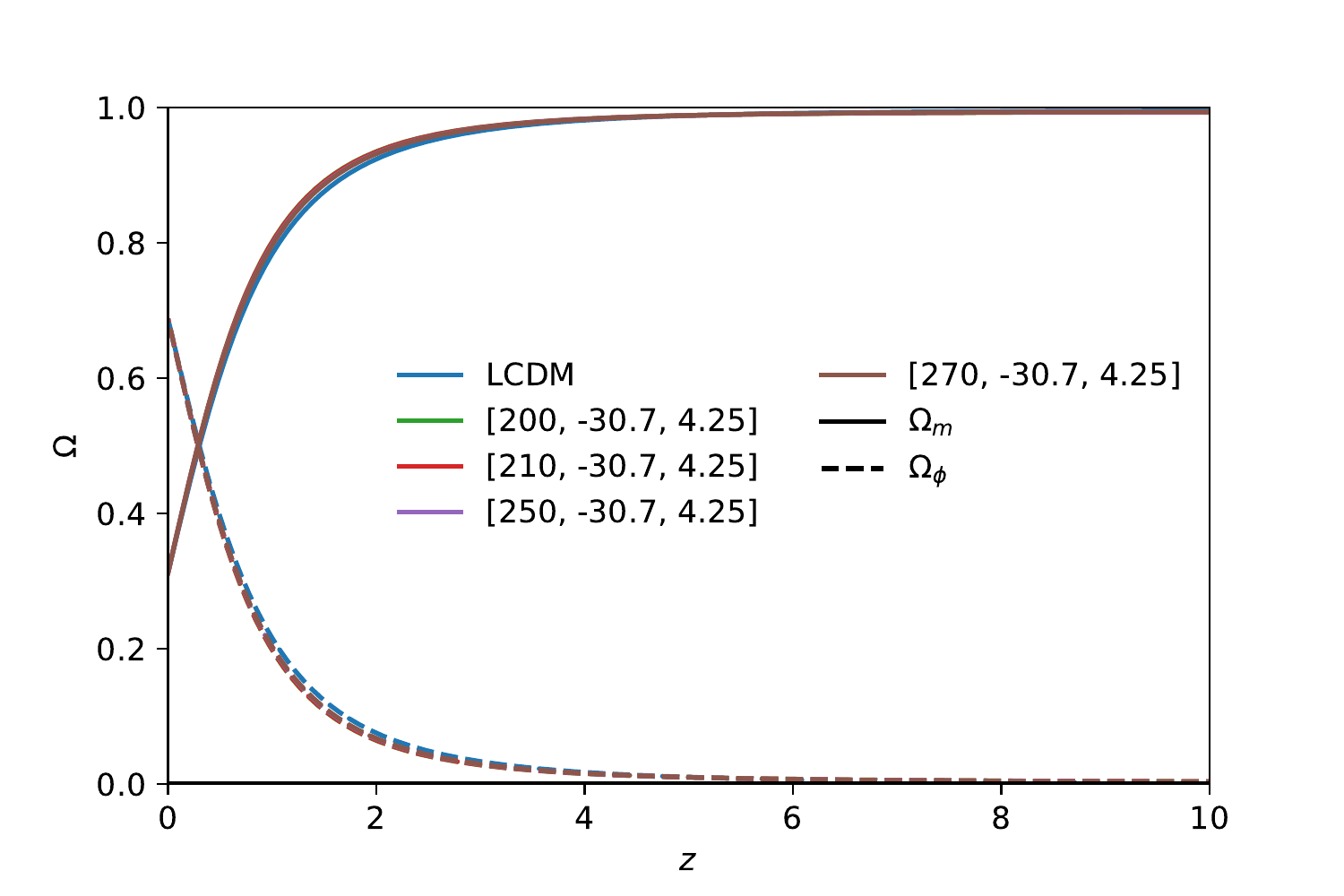}
\includegraphics[width=0.48\textwidth]{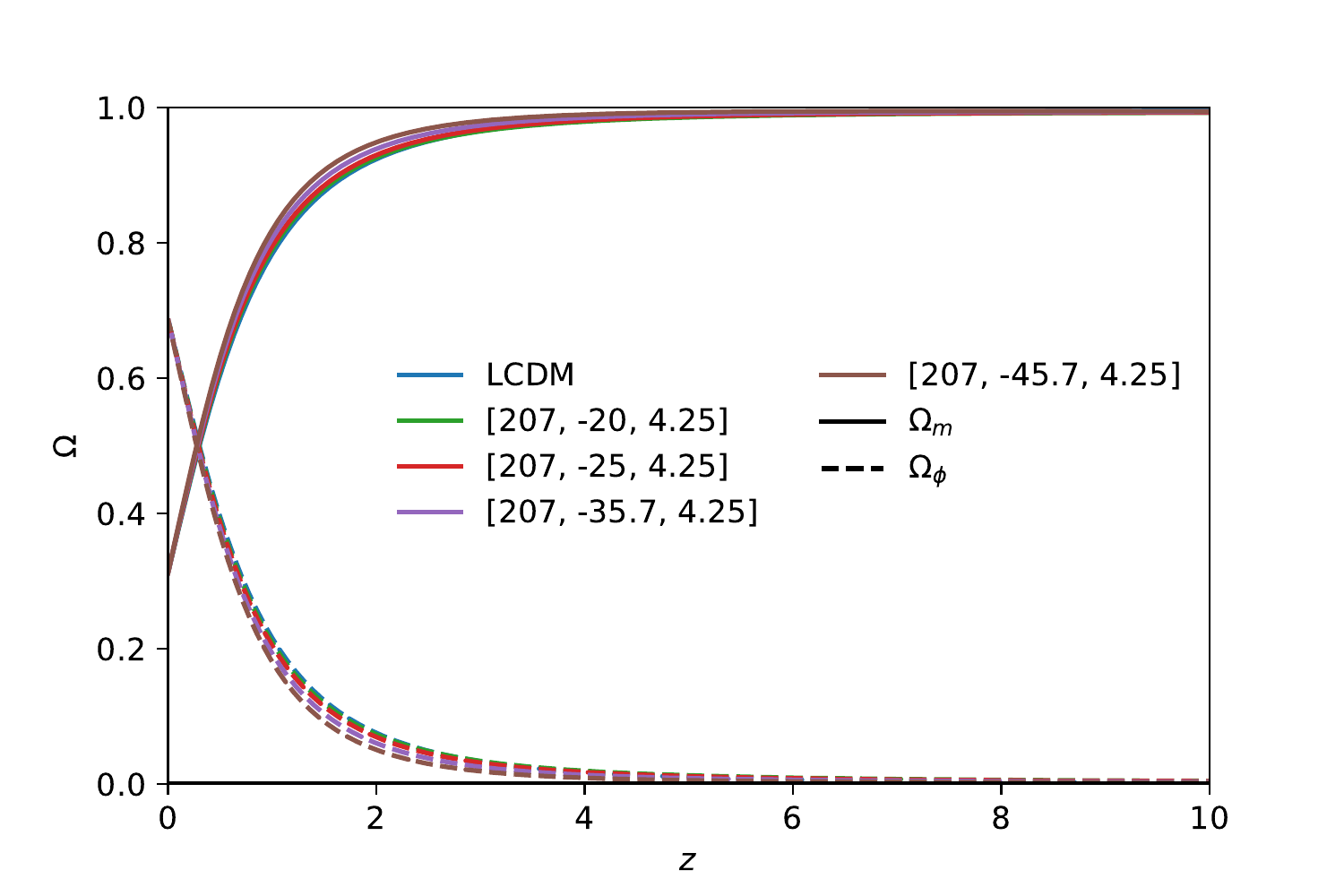}
\includegraphics[width=0.48\textwidth]{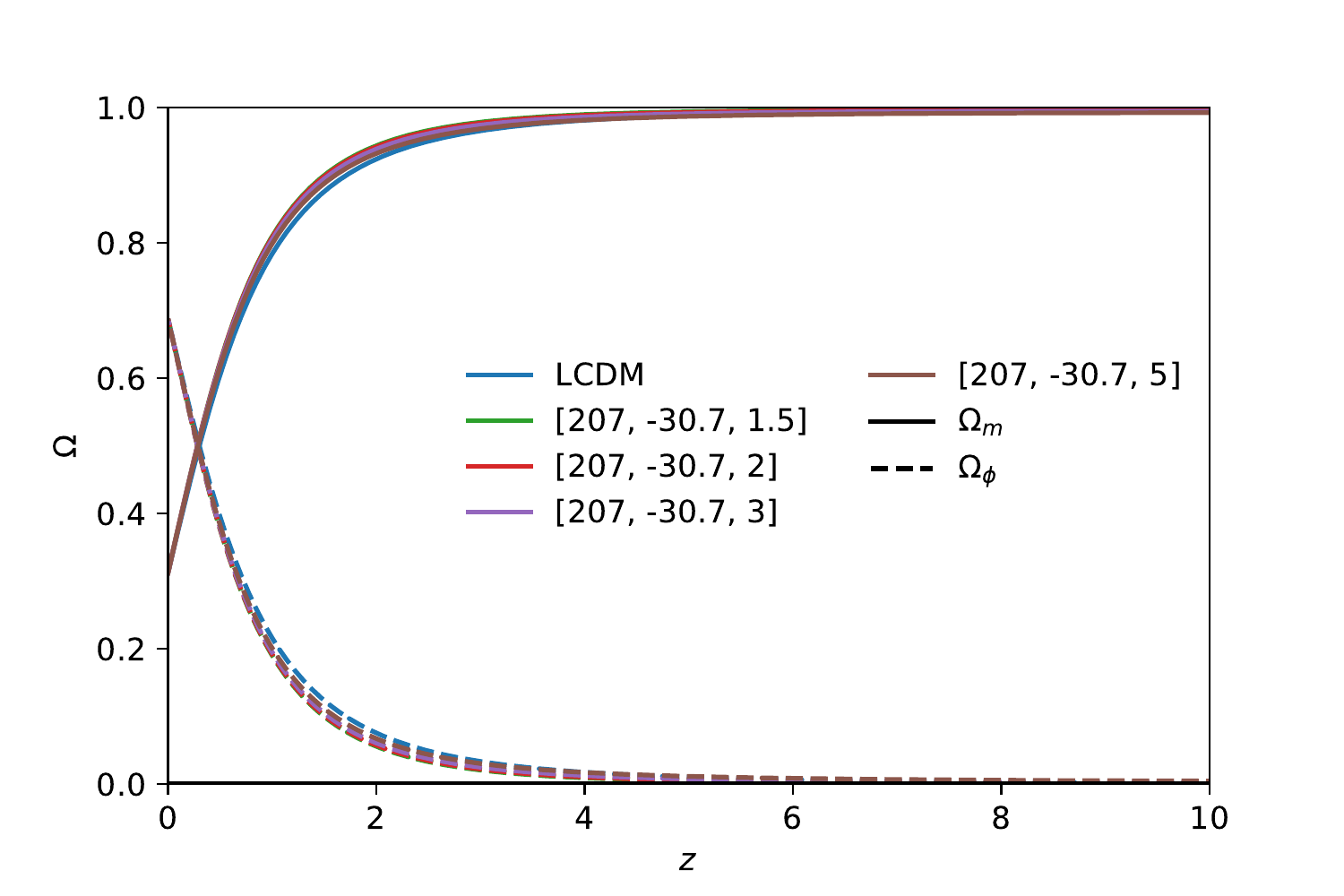}
\caption{\label{fig:density} The evolution of the dark energy density parameter ($\Omega_\phi (z)$) (dashed line) and the matter density parameter ($\Omega_m (z)$) (solid line). The top, middle and bottom panels correspond to the variation of the model parameters $w_0, w_1$ and $w_3$ respectively keeping other two respective model parameters fixed to the best fit values.}
\end{figure}

The evolution of the energy densities of the matter $\Omega_m$ and the scalar field $\Omega_\phi$ components has been shown in Fig \ref{fig:density}. We have considered the same set of parameter values for $w_0, w_1, w_3$ as in  Fig \ref{fig:EoS}. The blue curves represent the $\Lambda$CDM case for reference. It can be seen from the plots that the energy densities are not very sensitive to the model parameters $w_0, w_1, w_3$ though these parameters have significant effect on the evolution of the EoS of dark energy. 

\begin{figure}[h]
\includegraphics[width=0.48\textwidth]{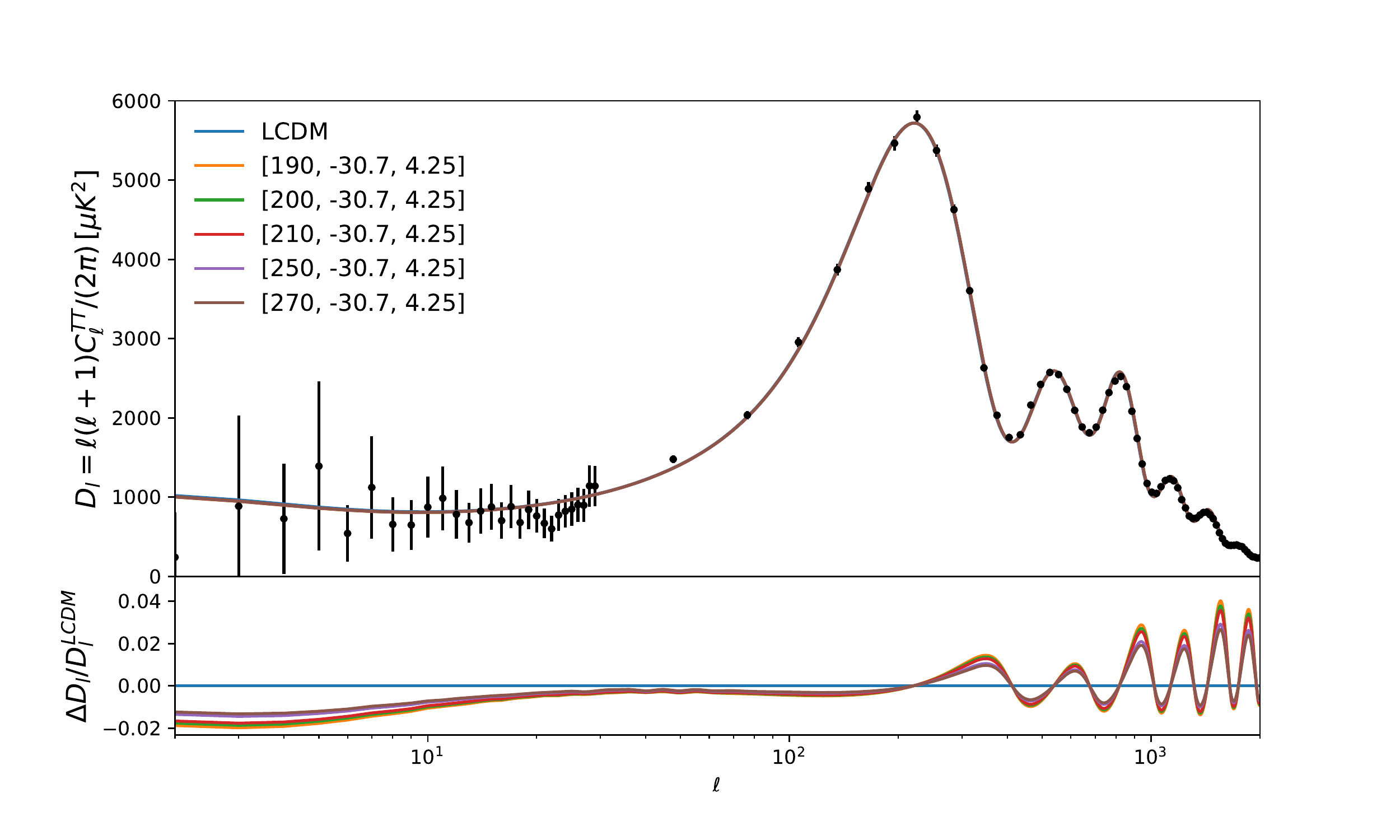}
\includegraphics[width=0.48\textwidth]{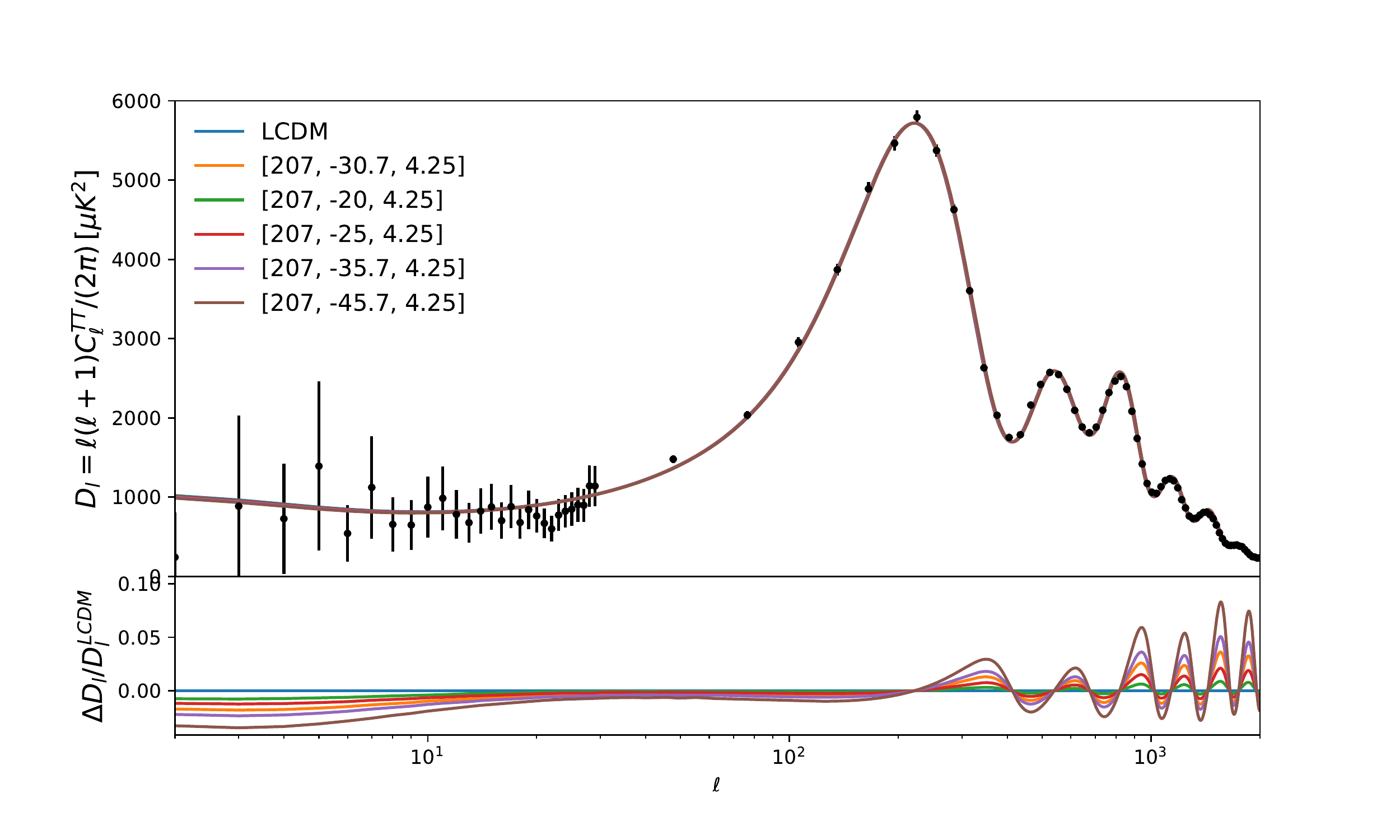}
\includegraphics[width=0.48\textwidth]{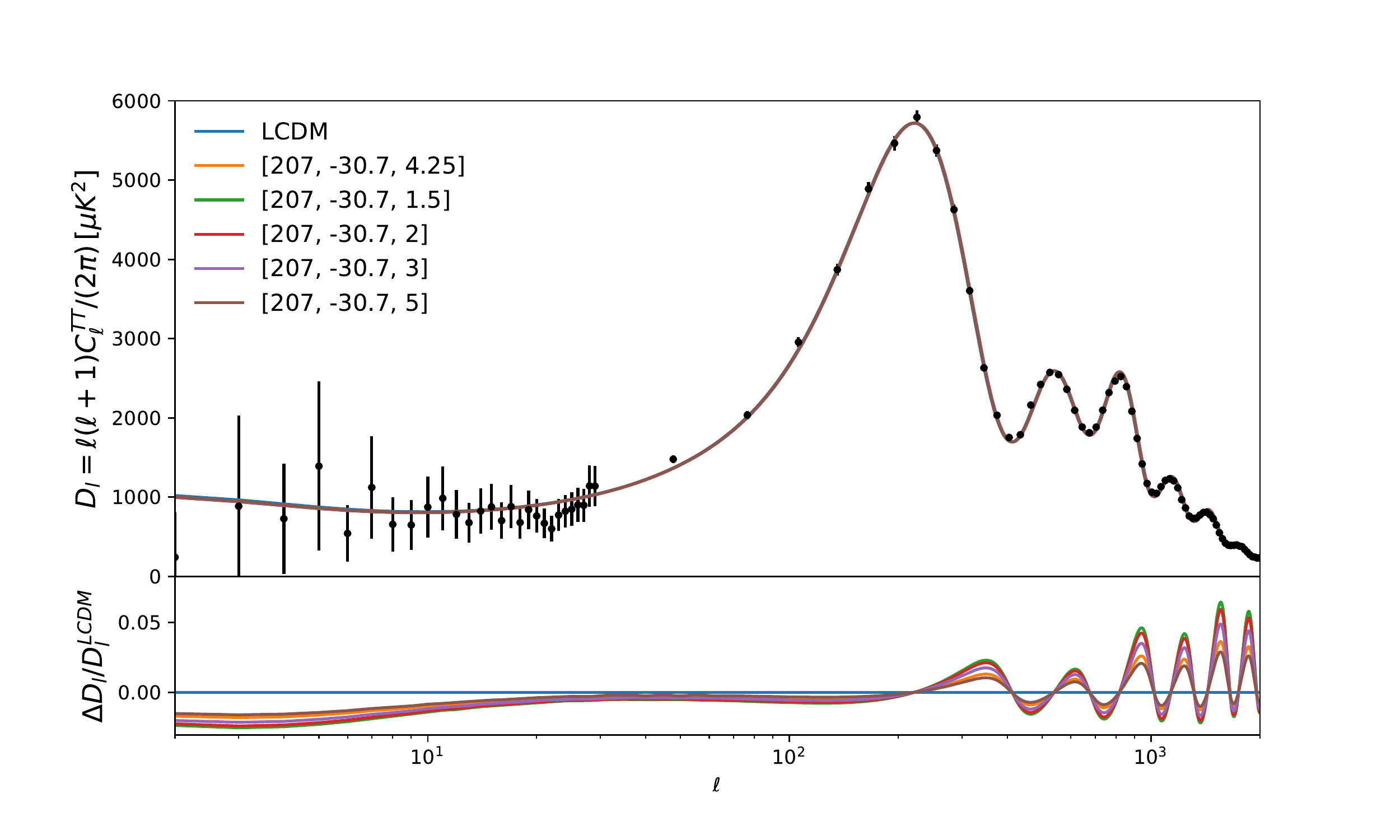}
\caption{\label{fig:tt} Plot of the CMB anisotropies for the $\phi CDM$ model. The top, middle and bottom panels correspond to the variation of the model parameters as explained in Fig \ref{fig:EoS}. The bottom panel of each figure shows the relative difference between the $\phi CDM$ and the $\Lambda$CDM ($ \Delta D_l =(D_l - D_l^{\Lambda CDM})/ D_l^{\Lambda CDM}$).}
\end{figure}

\begin{figure}[htb!]
\includegraphics[width=0.48\textwidth]{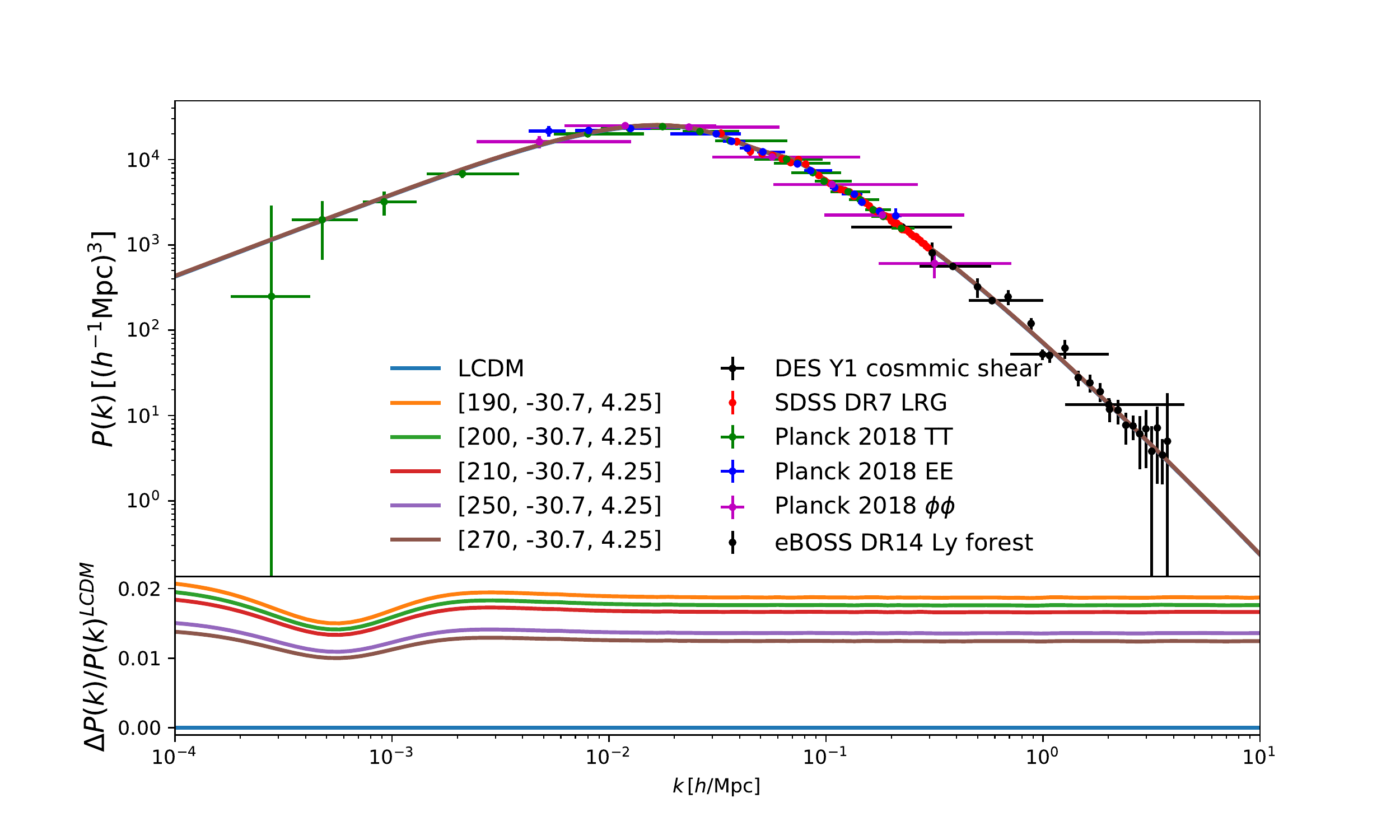}
\includegraphics[width=0.48\textwidth]{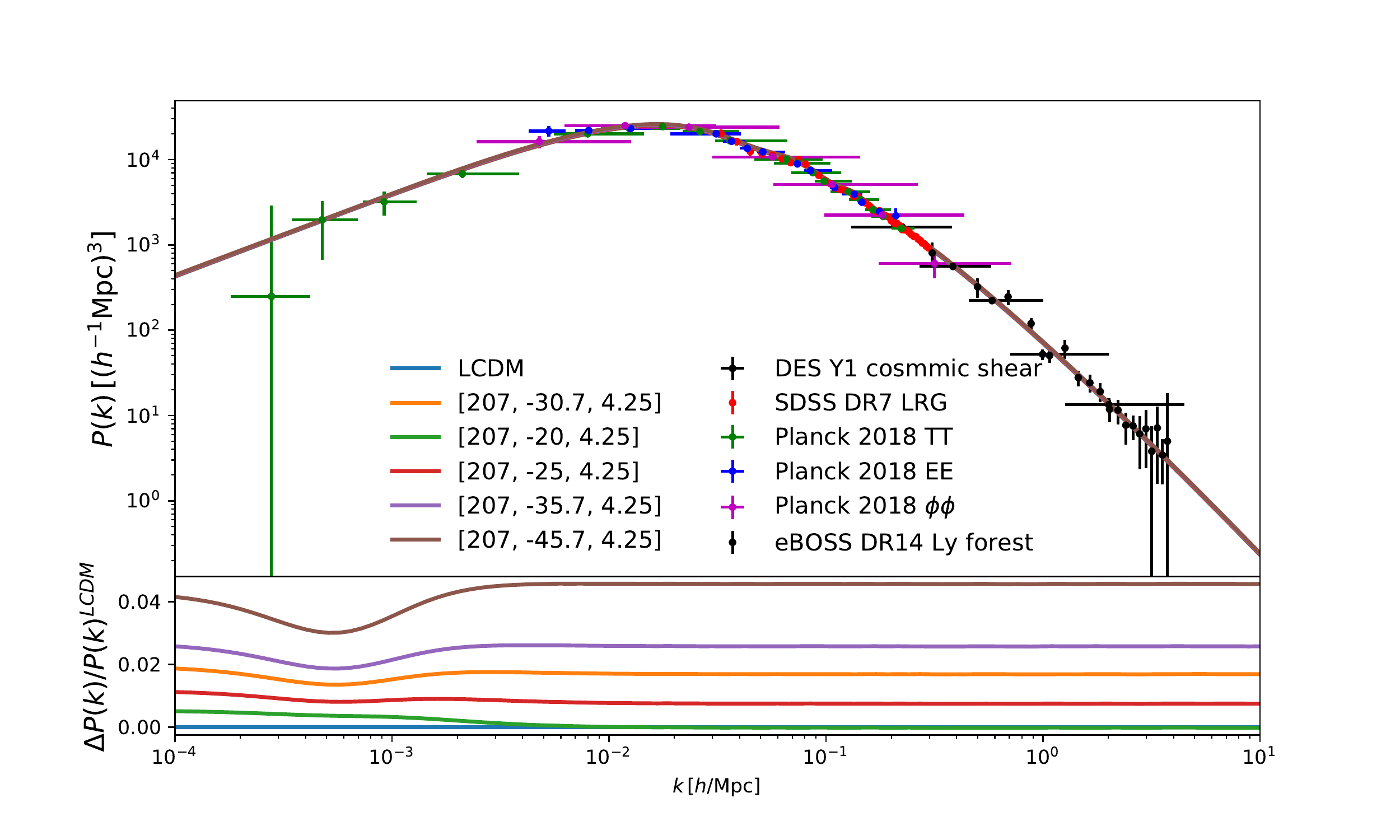}
\includegraphics[width=0.48\textwidth]{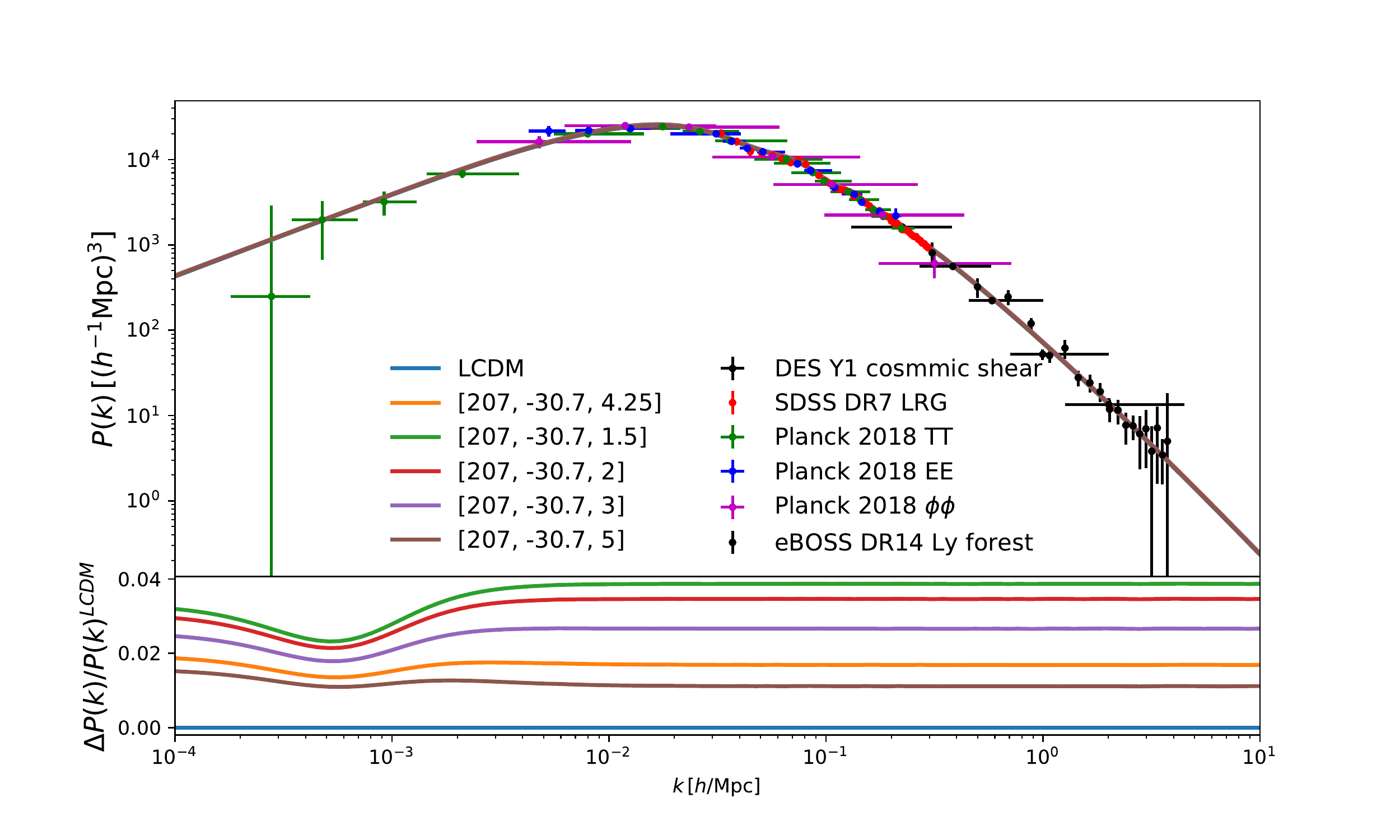}
\caption{\label{fig:mps} Top, middle and bottom panel shows the plots of matter power spectrum (MPS) for the variation of the model parameters as explained in Fig.\ref{fig:EoS}. At the bottom panel of each figure we showed relative difference between the $\phi CDM$ and the $\Lambda CDM$ models, ($ \Delta P(k) =(P(k) - P(k)^{\Lambda CDM})/ P(k)^{\Lambda CDM}$).}
\end{figure}

 We also show in  Fig \ref{fig:tt} and Fig \ref{fig:mps} the temperature anisotropies and the matter power spectrum (MPS) for the $\phi CDM$ model corresponding to the same set of model parameters as considered in Fig \ref{fig:EoS}. For comparison, data from different experiments have been plotted as references together with the numerical solutions. For the temperature anisotropies (TT) we have used binned TT power spectrum data from Planck 18 \cite{Planck:2019nip}.  For the MPS following data sets have been used: Planck2018 CMB data \cite{Planck:2019nip}, SDSS galaxy clustering \cite{Reid:2009xm}, SDSS Ly$\alpha$ forest \cite{SDSS:2017bih} and DES cosmic shear data \cite{DES:2017qwj} (for details on full data collection, please see  \cite{2019MNRAS.489.2247C}).
 
 The lower panels of Fig \ref{fig:tt} and Fig  \ref{fig:mps} represent the relative differences in $D_l$ and $P(k)$ in comparison to the $\Lambda CDM$ case. It is observed that in case of $D_l$, there is notable deviation from $\Lambda CDM$ case for both the lower and higher multipoles. For $P(k)$, the deviation from the $\Lambda CDM$ case is observed at all scales. Although the percentage deviation is not much, but one should, in principle, take into account the dark energy perturbations as well to obtain the complete  picture. A similar claim has been made in \cite{Urena-Lopez:2020npg} that if one does not consider the dark energy perturbations, this  can result in misleading constraints on the cosmological parameters.

\section{Conclusion}\label{conclusions}

In this work we have revisited the dynamics of the scalar field dark energy models  and proposed a general scheme which can include both the quintessence and the phantom scalar field models. Our method is simple and straightforward in which it is possible to express all the cosmological parameters in terms of the redshift ($z$), the present value of the matter density parameter ($\Omega_{m 0}$) and the normalized Hubble parameter ($E$). We have obtained the expressions for various cosmological parameters and have found that the final expressions of these cosmological parameters are independent of the nature of the scalar field, either quintessence or phantom. A general condition for the phantom barrier crossing of the dark energy models is also obtained. This general condition is particularly interesting as it can help us to check the possibility of phantom barrier crossing for a given DE model which complies with the methodology given here without a detailed numerical study of the model. This encourages us to consider a suitable parametrization of the Hubble parameter as toy model and to study the model against the current cosmological observations.

A publicly available version of the Boltzman code CLASS has been amended. The dynamics of the scalar field has been incorporated in the  CLASS code by adding it as a fluid using our proposed parametrization. A detailed study of the model against the recent cosmological data sets has been done using the MCMC code Montepython. The best fit value of the EoS parameter for the scalar field at the current epoch comes out to be less than $-1$ which indicates that at present phantom DE models are preferred by observational data compared to quintessence models. However a phantom barrier crossing has been observed within  $2\sigma$ confidence level for a wide range of choice of the model parameters. 

 The best fit value of $H_0$ comes out to be $68.9^{+0.573}_{-0.602}$  which happens to be more compatible with the PLANCK collaboration results and thus cannot alleviate the Hubble tension. This is in agreement with the recent results \cite{dinda2021cosmic} where it has been shown that CMB+BAO+SN data put stronger constraints on $H_0$ and on other background cosmological parameters and that the addition of $H_0$ prior from SH0ES (or from similar other local distance observations) can not significantly pull the $H_0$ value towards the corresponding SH0ES value. A comparison between $\phi CDM$ and $\Lambda CDM$ models have been carried out using the concept of Bayes Factor and the $\phi CDM$ model is found to have positive preference over the $\Lambda CDM$ one. We have also found that there are slight deviations in the $D_l$ and $P(k)$ curves as compared to standard $\Lambda CDM$; for $D_l$, deviation has been observed in both lower and higher multipoles whereas for $P(k)$ the deviation has been observed at all scales.  

We must admit that the choice of parametrization for $E(z)$ is motivated by mathematical simplicity and is not unique. A wide variety of parametrization can be considered which might be able to solve the $H_0$ tension as well. As of now, the current choice of parametrization valid for these two scalar field models can not alleviate the $H_0$ tension completely, but interesting results may be obtained by considering an interaction between the two dark sectors.

\section{Acknowledgement}
We are grateful to Ankan Mukherjee, Luis Urena Lopez and Francisco X. Linares Cede\~no for help in the making of some plots. We acknowledge the use of the Chalawan High Performance Computing cluster, operated and maintained by the National Astronomical Research Institute of Thailand (NARIT). The research of NR is supported by Mahidol University through the research project MRC-MGR 04/2565. SD would like to acknowledge IUCAA, Pune for providing support through the associateship programme. 

\appendix
\section{Cosmological parameters in terms of $w_0,w_1,w_2,w_3$}\label{appen:parameters}
In the text we have considered two different sets of parameters $[p, b, c, d]$ and $[w_0, w_1, w_2, w_3]$. Here we present all the cosmological parameters provided in equations (\ref{eq:om}) to (\ref{eq:vphi}) in terms of the parameters $w_0,w_1,w_2$ and $w_3$. 

 \begin{equation}\label{E_ana}
 E^2(z)=\frac{f_3}{f_2}
 \end{equation}

\begin{equation}
\begin{split}
q(z)=&\frac{1}{2f_3}  ( (f_1- w_2)(1+z)^3 + w_1(\Omega_{m0}-1)(1+z) \\
 &\quad +2 w_0 (\Omega_{m0}-1))
\end{split}
\end{equation}

\begin{equation}
\Omega_{m}(z)=\frac{\Omega_{m0}f_2(1+z)^3}{f_3}
\end{equation}

\begin{equation} \label{om_phi}
\Omega_{\phi}=1-\frac{\Omega_{m0}f_2(1+z)^3}{f_3} 
\end{equation}

\begin{equation}
\begin{split}
\frac{V(z)}{3H_{0}^2}=& \frac{1}{6 f_2}[ -(1+z)(3(f_1-w_2)(1+z)^2\\
&\quad -2w_3(\Omega_{m0}-1)(1+z)-w_1(\Omega_{m0}-1))\\
&\quad  +6(f_1 - w_2)(1+z)^3- 6 w_3 (\Omega_{m0}-1)(1+z)^2\\
&\quad -6 w_1 (\Omega_{m0}-1)(1+z) -6 w_0 (\Omega_{m0}-1)]\\
&\quad-\frac{1}{2}\Omega_{m0}(1+z)^3 
\end{split}
\end{equation}

\begin{align}
\phi(z)=\int \frac{1}{\epsilon(1+z) f_{3}}\left[\left(3(f_{1}-w_{2}\right)(1+z)^{2} \right.\nonumber\\
\left.\left.-2 w_{3}\left(\Omega_{m 0}-1\right)(1+z)\right.\right.\nonumber\\
\left.\left.-w_{1}\left(\Omega_{m 0}-1\right)\right)-3 \Omega_{m 0} f_{2}(1+z)^{2}\right]^{1 / 2} dz
\end{align}

where $$f_1 = (\Omega_{m0}(w_0+w_1+w_3)),$$  $$f_2 = (w_0+w_1+w_3-w_2)$$ and 

\begin{equation}
\begin{split}
f_3 = (f_1 - w_2)(1+z)^3- w_3 (\Omega_{m0}-1)(1+z)^2\\- w_1 (\Omega_{m0}-1)(1+z) - w_0 (\Omega_{m0}-1).\nonumber
\end{split}
\end{equation}

\section{Comparison between the analytical solution and the numerical solution} \label{appen:comparison}

\begin{figure}[h!]
\includegraphics[width=0.49\textwidth]{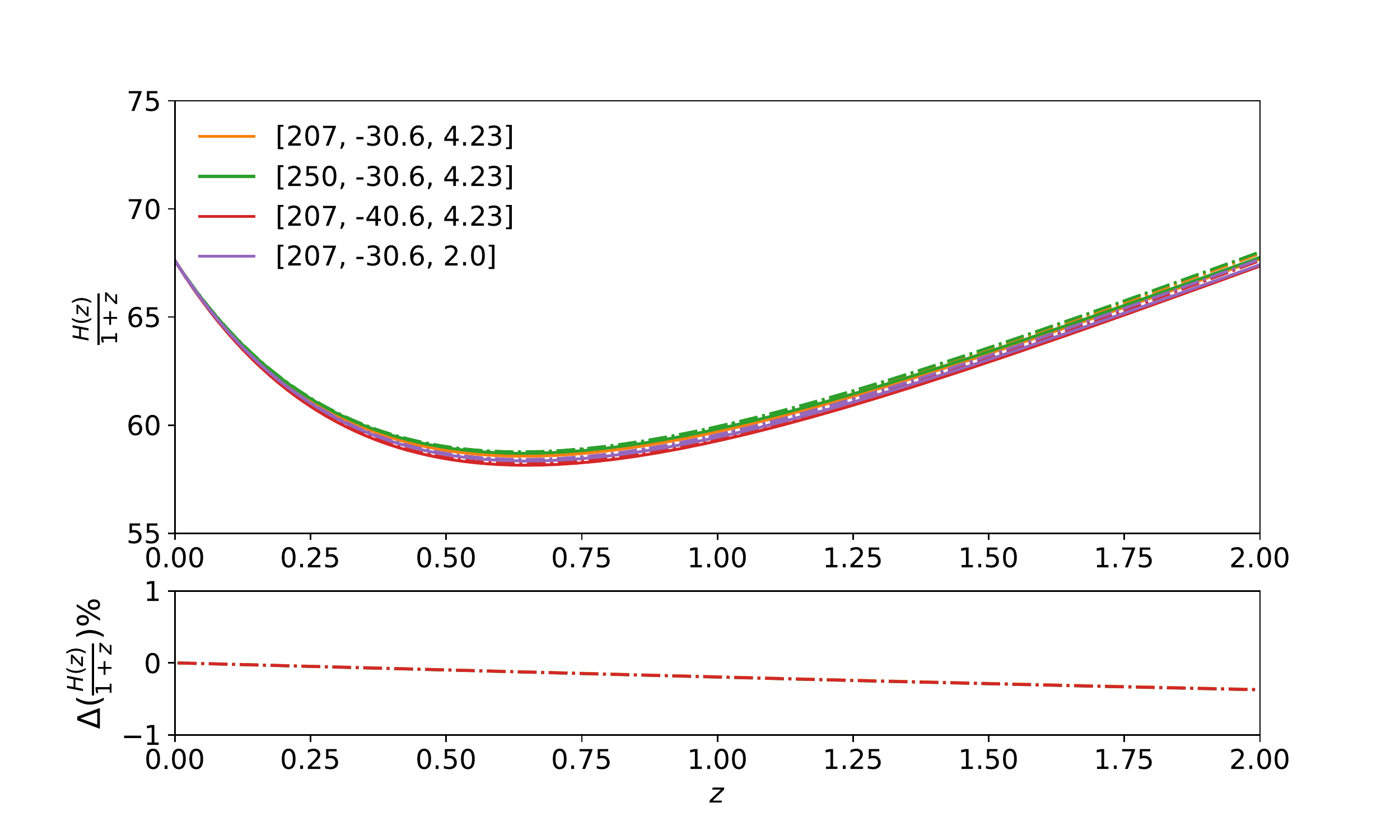}
\caption{\label{H_equi} Comparison of the ${H(z)}/{1+z}$ for the numerical solutions obtained from the CLASS code (represented by solid lines) and  the analytical solutions obtained from equation (\ref{E_ana}) (represented by dashed lines). In the bottom panel we have shown the percentage difference between the numerical solutions and the analytical solutions.}
\end{figure}

\begin{figure}[h!]
\includegraphics[width=0.49\textwidth]{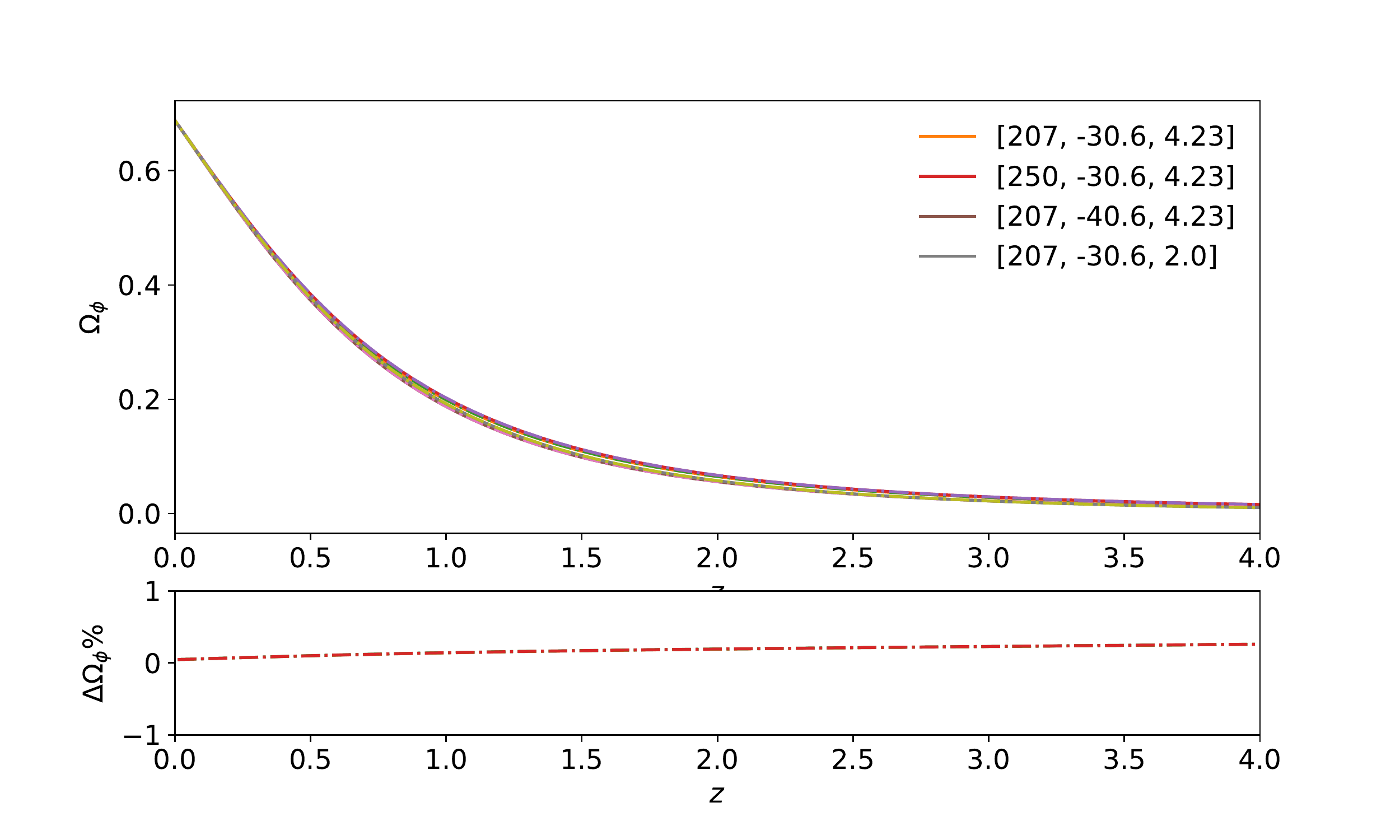}
\caption{\label{Omega_equi} Comparison of $\Omega_\phi (z)$ for the numerical solutions obtained from the CLASS code (represented by solid lines) and the analytical solutions obtained from equation (\ref{om_phi}) (represented by dashed lines). In the bottom panel we have shown the percentage difference between the two solutions.}
\end{figure}

 In Section \ref{sec:numerical} we have mentioned that  the dynamics of the scalar field has been incorporated in the CLASS code as a fluid by implementing the EoS of the DE given in equation (\ref{eq.EoS2}). Here we present a comparison between the numerical solutions obtained from the CLASS code and the analytical solutions of the cosmological parameters. In Fig \ref{H_equi} we have compared the expansion  rate (${H(z)}/{1+z}$) of the universe for these two solutions. The solid lines are the numerical solutions obtained from the CLASS code and the dashed lines are the corresponding plots for the analytical solutions which can be obtained from equation (\ref{E_ana}). The bottom panel of Fig \ref{H_equi} shows the percentage difference between the numerical and the analytical solutions. A similar comparison between the dark energy density $(\Omega_\phi)$ parameters for the numerical and the analytical solutions has been provided in the upper panel of Fig \ref{Omega_equi} along with the corresponding percentage difference at the bottom panel. One can see that for both the cases the percentage difference is less than $1\% $. It must be mentioned here that in order to obtain the analytical solutions, we have used the Klein Gordon (KG) equation for the scalar field whereas the numerical solutions obtained from the CLASS code involve the integral of the EoS of the dark energy over redshift interval. This similarity and very little deviation justify our claim in Section \ref{sec:numerical} that if one considers the EoS of the scalar field as a fluid in the CLASS code, it will resemble the same dynamics as that of the scalar field.

\newpage
\bibliography{qEoS}
\end{document}